\documentclass[%
 reprint,
superscriptaddress,
 amsmath,amssymb,
 aps,
prx,
]{revtex4-1}

\usepackage{graphicx}
\usepackage{float}
\usepackage{dcolumn}
\usepackage{bm}
\usepackage{booktabs}  
\usepackage{threeparttable}
\usepackage{multirow}
\usepackage{subfigure} 
\usepackage{mathtools}
\usepackage{algorithm}
\usepackage{algorithmic}
\usepackage{amsfonts,amssymb}
\usepackage{verbatim}

\begin{document}

\preprint{APS/123-QED}

\title{High-rate discretely-modulated continuous-variable quantum key distribution \\ using quantum machine learning}
\author{Qin Liao}
\email{llqqlq@hnu.edu.cn}
\affiliation{College of Computer Science and Electronic Engineering, Hunan University, Changsha 410082, China}

\author{Jieyu Liu}
\affiliation{College of Computer Science and Electronic Engineering, Hunan University, Changsha 410082, China}

\author{Anqi Huang}
\affiliation{Institute for Quantum Information \& State Key Laboratory of High Performance Computing, College of Computer Science and Technology, National University of Defense Technology, Changsha 410073, China}

\author{Lei Huang}
\affiliation{College of Computer Science and Electronic Engineering, Hunan University, Changsha 410082, China}

\author{Zhuoying Fei}
\affiliation{College of Computer Science and Electronic Engineering, Hunan University, Changsha 410082, China}

\author{Xiquan Fu}
\affiliation{College of Computer Science and Electronic Engineering, Hunan University, Changsha 410082, China}

\date{\today}

\begin{abstract}

We propose a high-rate scheme for discretely-modulated continuous-variable quantum key distribution (DM CVQKD) using quantum machine learning technologies, which divides the whole CVQKD system into three parts, i.e., the initialization part that is used for training and estimating quantum classiﬁer, the prediction part that is used for generating highly correlated raw keys, and the data-postprocessing part that generates the final secret key string shared by Alice and Bob. To this end, a low-complexity quantum $k$-nearest neighbor (Q$k$NN) classifier is designed for predicting the lossy discretely-modulated coherent states (DMCSs) at Bob's side. The performance of the proposed Q$k$NN-based CVQKD especially in terms of machine learning metrics and complexity is analyzed, and its theoretical security is proved by using semi-definite program (SDP) method. Numerical simulation shows that the secret key rate of our proposed scheme is explicitly superior to the existing DM CVQKD protocols, and it can be further enhanced with the increase of modulation variance. 

\begin{description}
\item[PACS numbers]
42.50.St
\end{description}
\end{abstract}

\pacs{Valid PACS appear here}
\maketitle


\section{\label{sec:level1} introduction}

Continuous-variable quantum key distribution \cite{Pirandola:20} is designed to implement point-to-point secret key distribution, its security is guaranteed by the fundamental laws of quantum physics \cite{r31}. In a basic version of CVQKD \cite{Grosshans:2002gm}, the sender, called Alice,  encodes secret key bits in the phase space of coherent states and sends them to an insecure quantum channel, while the receiver, called Bob, measures these incoming signal states with coherent detection. After several steps of data post-processing, a string of secret keys can be finally shared by Alice and Bob. One of the advantages of CVQKD is that it is compatible with most existing commercial telecommunication technologies \cite{r30,r32,Chen:2023}, making it easier to integrate into real-world communication links.

In general, a CVQKD system is mainly composed of quantum signal processing and data-postprocessing \cite{PhysRevLett.88.057902}. The former part corresponds to signal modulation, transmission, and measurement, aiming to generate a raw key, while the latter part corresponds to data reconciliation, parameter estimation, and privacy amplification, attempting to extract the final secret key from the raw key. In CVQKD, secret key rate and maximal transmission distance are generally a pair of crucial performance indicators. For a specific CVQKD system, however, there is tradeoff between the secret key rate and the maximal transmission distance: the longer the transmission distance, the lower the secret key rate, and vice versa. The main reason is that the continuous-variable quantum signal used to carry the secret key is extremely weak. Channel loss and excess noise rise as transmission distance increases, resulting in a reduction of signal-to-noise ratio (SNR) \cite{PhysRevA.76.042305}. This obliges a coherent detector hardly discriminate between the quantum signal and noise, decreasing the secret key rate. Although some solutions, such as adding an optical amplifier \cite{Fossier:2009dz} and adopting non-Gaussian operation \cite{Guo:2017ce,PhysRevA.93.012310}, can effectively improve the performance of the CVQKD system, its improvement seems limited as they are still largely constrained by the imperfect devices.

In recent years, CVQKD using machine learning technologies is becoming a research hotspot, as these technologies can be used for improving the performance of CVQKD without any extra device \cite{PhysRevA.106.022607}. More importantly, machine learning technologies can automatically compensate for the negative effects caused by imperfect devices, effectively removing the performance restriction of the imperfect devices. For instance, Ref.~\cite{Liao2018Long} reported a state-discrimination detector based on a Bayesian classification algorithm. This detector has the ability to surpass the standard quantum limit, which can be only achieved by conventional ideal detectors \cite{Becerra:2013jw}, so that the maximum transmission distance of the CVQKD system can be significantly increased. Reference \cite{Liao_2020} suggested a multi-label learning-based embedded classifier with the capability to precisely predict the location of the signal state in phase space, so that it can dramatically improve the performance of the CVQKD system as well. 

Although these works do reveal that machine learning-based technologies can significantly improve the performance of CVQKD system, an underlying issue seems to be neglected. With an extremely large amount of signal data, the complexity of the majority of machine learning technologies is nearly unacceptable, and this situation is especially severe in high-speed CVQKD system \cite{Pirandola2015High}. For example, the core machine learning technology used in Ref.~\cite{Liao_2020} is based on $k$-nearest neighbor ($k$NN) \cite{1053964}, which is one of the most mature classification algorithms. However, $k$NN has a very high complexity as each unlabeled instance has to calculate all the distances between all labeled instances and itself, rendering extraordinary consumption in both time and space. These heavy costs have to be well addressed, otherwise CVQKD using machine learning technologies is unpractical, especially for the high-speed and/or real-time secret key distribution scenarios.

In past few years, quantum machine learning, which is based on quantum algorithms such as Shor's algorithm \cite{RevModPhys.68.733} and Grover's algorithm \cite{grover1996fast}, has been developed rapidly. For example, Ref.~\cite{PhysRevLett.109.050505} proposed a quantum linear regression algorithm based on the quantum HHL algorithm \cite{PhysRevLett.103.150502}, which exponentially accelerates the classical algorithm. Reference \cite{PhysRevLett.113.130503} reported a quantum support vector machine (SVM) which used the high parallelism of quantum computing to improve the classical SVM, thus obtained an exponential speedup. References \cite{aimeur2007quantum, aimeur2013quantum} showed that the classical $K$-means algorithm can be accelerated by quantum minimal finding method. The above studies indicate that quantum algorithms contribute to speed up machine learning, thereby improving the computing efficiency. 

Inspired by the merits of quantum algorithms, in this work, we propose a high-rate CVQKD scheme based on quantum machine learning. The proposed scheme is quite different from conventional discretely-modulated (DM) CVQKD \cite{PhysRevLett.102.180504, PhysRevA.83.042312}, which divides the whole CVQKD system into three parts, initialization, prediction and data-postprocessing. The initialization part is used for training and estimating quantum classiﬁer, the prediction part is used for generating highly correlated raw keys, and the data-postprocessing part generates the final secret key string shared by Alice and Bob. In particular, a well-behaved quantum $k$-nearest neighbor (Q$k$NN) algorithm is designed as a quantum classifier for distinguishing the lossy discretely-modulated coherent states (DMCSs) at Bob's side. Different from classical $k$NN algorithm, Q$k$NN simultaneously calculates all similarities in parallel and sorts them by taking advantages of Grover's algorithm, thereby greatly reducing the complexity. The performance of the proposed Q$k$NN-based CVQKD especially in terms of machine learning metrics and complexity is analyzed, and its theoretical security is proved by using semi-definite program (SDP) method. Numerical simulation shows that the secret key rate of Q$k$NN-based CVQKD is explicitly superior to the existing DM CVQKD protocols, and it can be further enhanced with the increase of modulation variance. It indicates that the proposed Q$k$NN-based CVQKD is suitable for the high-speed metropolitan secure communication due to its advantages of high-rate and low-complexity.

This paper is structured as follows. In Sec.~\uppercase\expandafter{\romannumeral2}, we briefly introduce the CVQKD protocol and classical $k$NN algorithm. In Sec.~\uppercase\expandafter{\romannumeral3}, we detail the proposed Q$k$NN-based CVQKD. Performance analysis and discussion are presented in Sec.~\uppercase\expandafter{\romannumeral4} and conclusions are drawn in Sec.~\uppercase\expandafter{\romannumeral5}.

\section{CVQKD protocol and classical $k$NN algorithm}

In order to make the paper self-contained, in this section, we first retrospect the process of CVQKD protocol, and briefly introduce the classical $k$NN algorithm.

\subsection{Process of CVQKD protocol}

\begin{figure}
\centering
\includegraphics[scale=0.4]{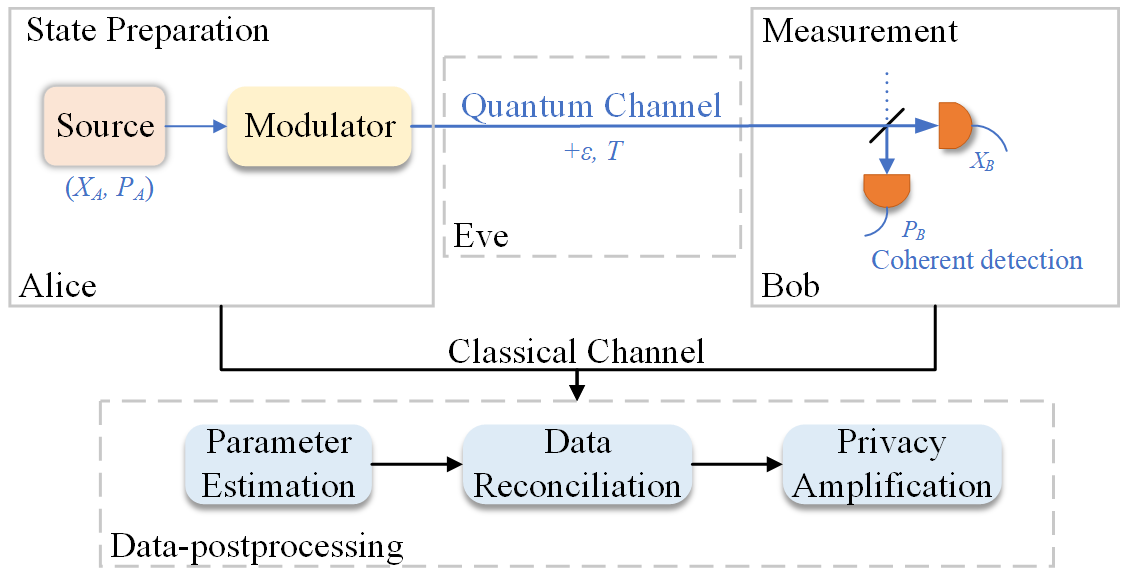}
\caption{The process of conventional CVQKD. Alice ﬁrst prepares a modulated coherent state, and sends it to Bob through an untrusted quantum channel. Bob measures the incoming state with coherent detection so that Alice and Bob share two correlated sets variables, i.e. raw key. The final secret key can be obtained after data-postprocessing, which includes parameter estimation, data reconciliation and privacy amplification.}
\label{fig:CVQKD}
\end{figure}

In general, the whole process of one-way CVQKD protocol includes five steps, i.e., state preparation, measurement, parameter estimation, data reconciliation and privacy amplification. Figure \ref{fig:CVQKD} shows the process of conventional CVQKD, and each step is explained as follows.

{\bf State Preparation.} Alice prepares a train of coherent states whose quadrature values $x$ and $p$ obey a bivariate Gaussian distribution. Then the modulated coherent states are sent to Bob through an untrusted quantum channel.

{\bf Measurement.} The pulses sent by Alice are received by Bob who measures these incoming signal states with coherent detector. After enough rounds, a string of raw key can be shared between Alice and Bob.

{\bf Parameter Estimation.} Alice and Bob disclose part of the raw key to calculate the transmittance $T$ and excess noise $\varepsilon$ of the quantum channel. With these two parameters, the secret key rate can be estimated.

{\bf Data Reconciliation.} If the estimated secret key rate is positive, the low density parity check (LDPC) code is applied to another part of the raw key, aiming to error correction.  

{\bf Privacy Amplification.} Finally, a privacy amplification algorithm is performed based on the hash function, so as to extract the final secret key that is entirely unknown to the eavesdropper Eve.

\subsection{Classical $k$NN algorithm}\label{2.1}

\begin{figure}
\centering
\includegraphics[scale=0.65]{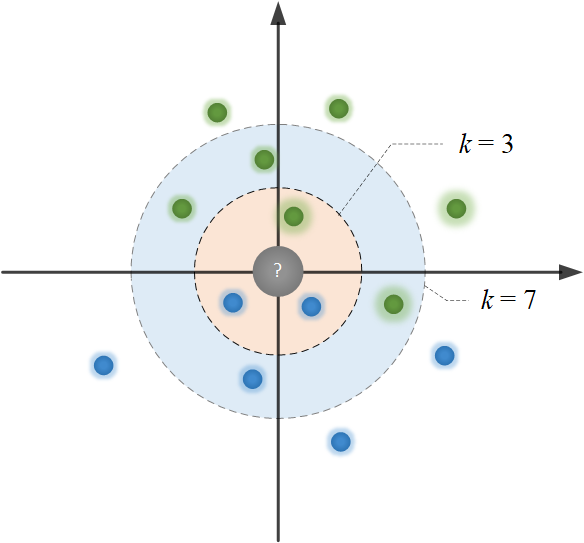}
\caption{An example of $k$NN algorithm in 2-dimensional feature space. The gray dot denotes an unlabeled data point. The green dots and blue dots are two classes of labeled data points. The gray dot will be assigned to the blue class when $k$=3, and it will be assigned to the green class when $k$=7.}
\label{fig:kNN}
\end{figure}

$k$NN is a traditional lazy learning algorithm that uses a majority vote to classify (predict) the grouping of unlabeled data points. For an unlabeled data point $v_0$ and a training set containing $v_j \; (j=1,2,...,M)$ labeled data points, $k$NN first finds a set of labeled data points whose similarities with $v_0$ are the top $k$ highest among all data pints in training set, and then counts the number of each label, the label with biggest number will be assigned as the label of data point $v_0$. Figure \ref{fig:kNN} depicts an example of $k$NN algorithm in 2-dimensional feature space, in which a gray dot denotes an unlabeled data point and it is going to be labeled as green or blue. The gray dot will be assigned to the blue class when $k=3$ due to there are two of the three-nearest labeled data points belong to the blue class while only one point belongs to the green class. Similarly, it will be labeled as green class for $k=7$ as there are four of the seven-nearest labeled data points belong to the green class while only three point belongs to the blue class.

The specific steps of $k$NN can be seen in Table \ref{tab:knn}. It is worth noting that the similarity is one of the crucial parameters of $k$NN algorithm. Assuming ${\bf v_0}=(v_{01},v_{02},...,v_{0U})$ and ${\bf v_j}=(v_{j1},v_{j2},...,v_{jU})$ are the respective feature vectors of unlabeled data point $v_0$ and labeled data point $v_j$, similarity can be measured by following ways. 

{\bf Euclidean distance}
\begin{equation}
{\rm Sim}_E({\bf v_j},{\bf v_0})=\sqrt{\sum_{r=1}^U(v_{jr}-v_{0r})^2}.
\end{equation}

{\bf Cosine similarity}
\begin{equation}
{\rm Sim}_C({\bf v_j},{\bf v_0})=\frac{{\bf v_j}\cdot {\bf v_0}}{|{\bf v_j}||{\bf v_0}|}=\frac{\sum_{r=1}^U v_{jr}v_{0r}}{\sqrt{\sum_{r=1}^U v_{jr}^2}\sqrt{\sum_{r=1}^U v_{0r}^2}}.
\end{equation}

In the above two similarities, Euclidean distance counts the absolute distance of each data point, revealing the difference of each data point's location in the coordinate system \cite{van2005classification}. Cosine similarity counts the cosine of an angle between two vectors, revealing the directional difference of each vector \cite{tan2016introduction}. In addition, Manhattan distance \cite{stigler1986history}, Hamming distance \cite{waggener1995pulse} and inner product also can be used as similarity in $k$NN, one should select a proper way according to different applications.

\begin{table}
\caption{$k$NN Algorithm.}
\begin{ruledtabular}\label{tab:knn}
\begin{tabular}{lcr}
{\bf Input} : training data points $V_j=\{{\bf v_1},{\bf v_2},...,{\bf v_M}\}$ and their \\
labels $V_j^L=\{v_1^L,v_2^L,...,v_M^L\}$, unlabeled data point ${\bf v_0}$ and \\
hyper-parameter $k$.\\
{\bf Output} : predicted label $v_0^L$ of data point ${\bf v_0}$.\\
\hline
{\bf function} Predict\\
\qquad Load all training data points $V_j$ and unlabeled data \\ point ${\bf v_0}$ on the register.\\
\qquad{\bf for} $j\gets1$ to $M$ {\bf do}\\
\qquad\qquad Compute ${\rm Sim}_{0j}=S({\bf v_0},{\bf v_j})$.\\
\qquad{\bf end for}\\
\qquad Sort $S=\{{\rm Sim}_{01},{\rm Sim}_{02},...{\rm Sim}_{0M}\}$ (descending or \\
ascending).\\
\qquad Choose $k$ neighbors which are nearest to ${\bf v_0}$.\\
\qquad Conduct majority voting and assign the label $v_j^L$ of \\ the majority to the data point ${\bf v_0}$.\\
\qquad{\bf return} $v_0^L$\\
{\bf end function}
\end{tabular}
\end{ruledtabular}
\end{table}

\section{Q$k$NN-based CVQKD}\label{sec3}

\begin{figure}
\centering
\centering
\includegraphics[scale=0.5]{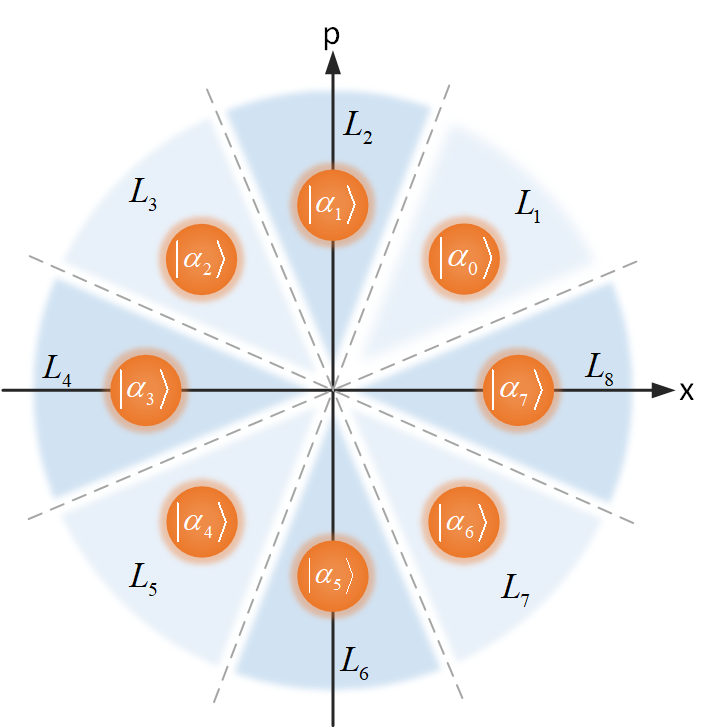}
\caption{Phase space representation of coherent states with 8PSK modulation. The orange dots $|\alpha_0\rangle$ to $|\alpha_7\rangle$ are standard 8PSK-modulated coherent states. Each coherent state is assigned to a label according to its location.}
\label{fig:label}
\end{figure}

\begin{figure*}
\centering
\includegraphics[scale=0.45]{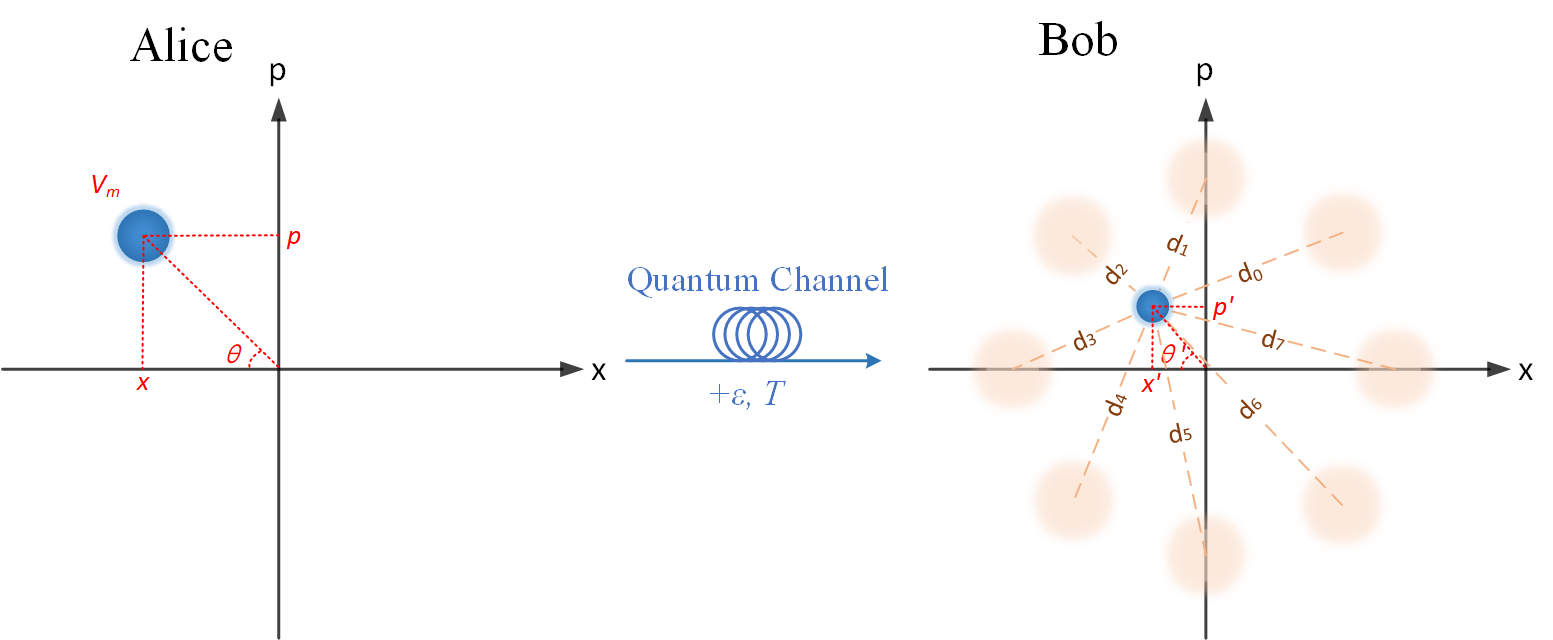}
\caption{Feature construction for coherent state in phase space. (Left) Alice randomly selects one of the 8PSK-modulated coherent states and sends it to the untrusted quantum channel. (Right) Bob receives the transmitted coherent state and extracts its eight-dimensional distance features $(d_0,d_1,d_2,...,d_7)$.}
\label{fig:feature}
\end{figure*}

Before we detail the proposed Q$k$NN-based CVQKD, several concepts need to be introduced. As shown in Fig.~\ref{fig:label}, the coherent states that Alice prepared are discretely modulated with 8 phase shift keying (8PSK), we equally divide the entire phase space into eight areas, and assign each area to a label $L_i \; (i=1,2,...,8)$. By doing so, each coherent state can be classified into a certain label according to its position. For example, coherent state $|\alpha_0\rangle$ belongs to label $L_1$ and coherent state $|\alpha_7\rangle$ belongs to label $L_8$. Note that although our study is mainly based on 8PSK modulation strategy, other discrete modulation strategies can also be used for the proposed Q$k$NN-based CVQKD. Figure \ref{fig:feature} shows the change of location in phase space of a modulated coherent state after passing an untrusted quantum channel. It depicts that the transmitted coherent state received by Bob is no longer identical with its initial signal state due to the phase drift $(\theta'\neq\theta)$ and energy attenuation $(\sqrt{x'^2+p'^2}<\sqrt{x^2+p^2})$ caused by the imperfect channel noise and loss. To detailedly describe this difference, we construct a multi-dimensional feature vector for each received coherent state by calculating Euclidean distances between the received coherent state and each standard 8PSK-modulated coherent state. By extracting these distance features, the features of received coherent state can be extended from three dimensions $(x', p', \theta')$ to eight dimensions $(d_0,d_1,d_2,...,d_7)$. For now, the Q$k$NN-based CVQKD can be detailed below.

\begin{figure*}
\centering
\includegraphics[scale=0.37]{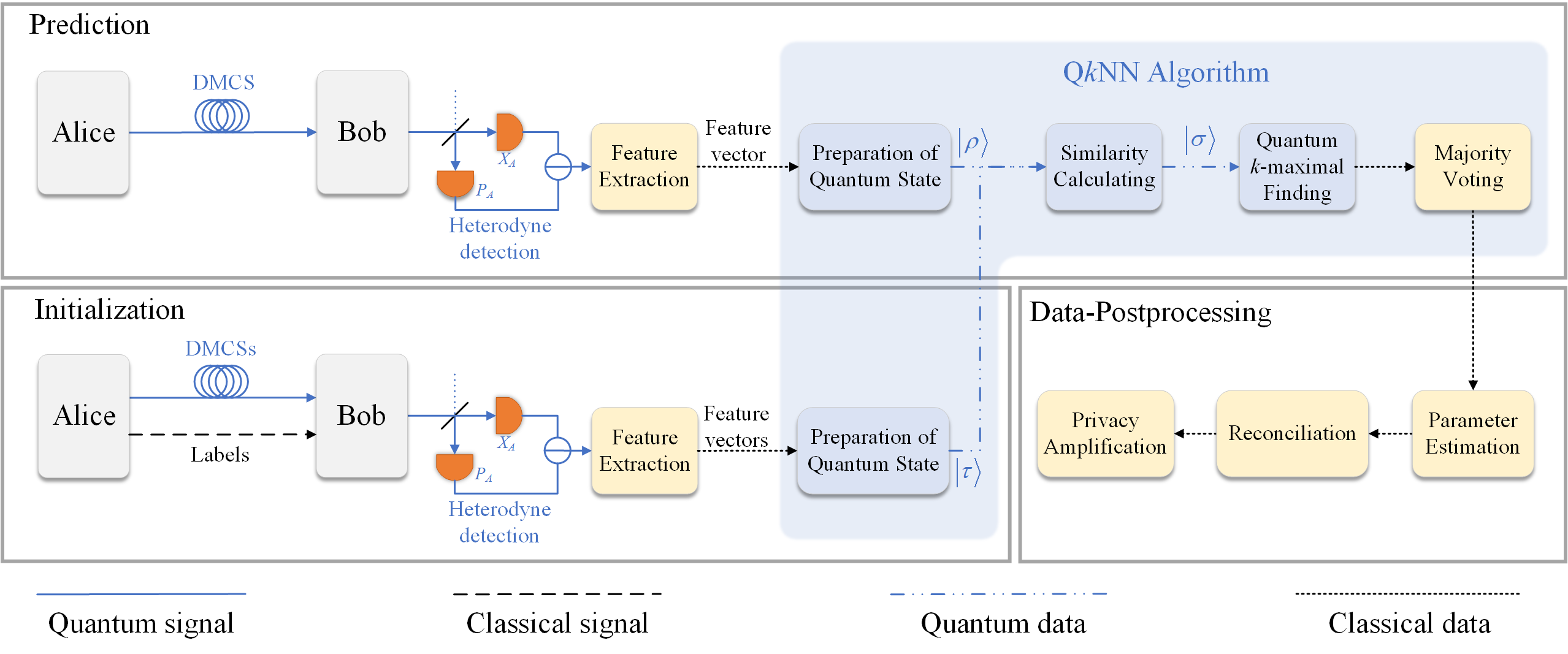}
\caption{The process of Q$k$NN-based CVQKD is divided into initialization, prediction and data-postprocessing. Yellow boxes represent classical operation, and blue boxes represent quantum operation.}
\label{fig:QkNN_QKD}
\end{figure*}

As shown in Fig.~\ref{fig:QkNN_QKD}, the whole Q$k$NN-based CVQKD system can be divided into three parts, i.e., initialization, prediction and data-postprocessing. 

\textbf{Initialization.} Alice first prepares DMCSs with 8PSK modulation and labels each DMCS which follows the rule shown in Fig.~\ref{fig:label}. Both DMCSs and their respective labels are sent to Bob who measures the received signals with heterodyne detection, and these DMCSs are called training data. Bob subsequently extracts features from each measurement result to obtain an eight-dimensional feature vector. After collecting enough feature vectors, Bob locally prepares a quantum state $|\tau\rangle$ to carry the information of all feature vectors. Note that the aim of the initialization part is to prepare quantum state $|\tau\rangle$ which stores feature information of all labeled DMCSs, so that once the quantum state $|\tau\rangle$ has been successfully prepared, one does not need to perform this part repeatedly. To guarantee the security, the transmission of labels has to be done without eavesdropping. This can be implemented by multiple ways, such as security monitoring of the communication system when the information of labels is transmitting. Even if the information of labels is compromised, the initialization part can be rerun to prepare new DMCSs and their respective labels.

\textbf{Prediction.} Alice randomly prepares an unlabeled DMCS with 8PSK modulation and sent it to the untrusted quantum channel. Bob receives this incoming signal state and measures it with heterodyne detection. Similarly, an eight-dimensional feature vector for this signal state can be constructed according to its measurement result. Bob then locally prepares a quantum state $|\rho\rangle$ to carry the information of this feature vector. For now, Bob holds both quantum state $|\rho\rangle$ that stores the information of unlabeled DMCS and quantum state $|\tau\rangle$ that stores the information of labeled DMCSs. He subsequently calculates the similarities between these quantum states and stores them into quantum state $|\sigma\rangle$. After quantum $k$-maximal finding, $k$ nearest neighbors of the unlabeled DMCS can be found. Finally, the label whose count is the biggest in $k$ nearest neighbors is assigned to the unlabeled DMCS, so that Bob can precisely recover the bit information sent by Alice. After enough rounds, Alice and Bob can share a string of raw key. Compared with the raw key generated by conventional DM CVQKD, the raw key generated by this part is more correlated, as quantum machine learning is introduced for better correcting the bit error caused by channel loss, excess noise and quantum attacks during the signal transmission.

\textbf{Data-Postprocessing.} This part is similar to the postprocessing of conventional CVQKD, which includes parameter estimation, data reconciliation and privacy amplification. Details about these steps can be found in Ref.~\cite{Leverrier:2015he}.

With these three parts, secret key can be finally shared between Alice and Bob. In what follows, several critical steps of the proposed Q$k$NN-based CVQKD is detailed.

\subsection{Preparation of quantum states}\label{sec3a}

As we mentioned above, Bob needs to prepare quantum states ($|\tau\rangle$ or $|\rho\rangle$) in both initialization part and quantum computing processing part. The difference is that quantum state $|\tau\rangle$ is prepared in initialization part to carry the information of all feature vectors that extracted from labeled DMCSs, while quantum states $|\rho\rangle$ is prepared in prediction part to carry the information of feature vector that extracted from an unlabeled DMCS. Assuming the normalized feature vectors that extracted from labeled DMCSs are $V=\{{\bf v_1},{\bf v_2},...,{\bf v_M}\}$ and the normalized feature vector that extracted from an unlabeled DMCS is ${\bf v_0}$, the quantum states $|\tau\rangle$ and $|\rho\rangle$ can be respectively expressed by \cite{wiebe2014quantum}
\begin{equation}\label{eq:tau}
\begin{aligned}
|\tau\rangle =\frac{1}{\sqrt{M}}\sum_{j=1}^M|j\rangle \frac{1}{\sqrt{U}}\sum_{i=1}^U|i\rangle|1\rangle(\sqrt{1-v_{ji}^2}|0\rangle+v_{ji}|1\rangle),
\end{aligned}
\end{equation}
and
\begin{equation}
\begin{aligned}
|\rho\rangle =\frac{1}{\sqrt{U}}\sum_{i=1}^U|i\rangle(\sqrt{1-v_{0i}^2}|0\rangle+v_{0i}|1\rangle)|1\rangle
\end{aligned}
\end{equation}
where $v_{ji}$ $(v_{0i})$ is the $i$-th feature value of feature vector ${\bf v_j}$ $({\bf v_0})$, $U$ is the dimension of the extracted features ($U=8$ in our case) and $M$ is the number of labeled DMCSs. Obviously, $|\tau\rangle$ is the superposition state that carries the information of all feature vectors for known (labeled) DMCSs, while $|\rho\rangle$ is the quantum state that carries the information of feature vector for an unknown (unlabeled) DMCS.

In what follows, we show how to obtain these two quantum states. To prepare quantum state $|\tau\rangle$, we first need to prepare an initial state $|\tau_{init}\rangle$ which can be expressed as
\begin{equation}
\begin{aligned}
|\tau_{init}\rangle=\frac{1}{\sqrt{M}}\sum_{j=1}^M|j\rangle \frac{1}{\sqrt{U}}\sum_{i=1}^U|i\rangle|0\rangle|0\rangle,
\end{aligned}
\end{equation}
where state $\frac{1}{\sqrt{M}}\sum_{j=1}^M|j\rangle$ (or state $\frac{1}{\sqrt{U}}\sum_{i=1}^U|i\rangle$) can be obtained by a quantum circuit shown in Fig.~\ref{fig:inintialstate}. See Appendix \ref{SP} for the detailed derivation of initial state $|\tau_{init}\rangle$. Subsequently, an Oracle ${\mathcal O}|j\rangle|i\rangle|0\rangle=|j\rangle|i\rangle|v_{ji}\rangle$ is applied to this initial state so that the resultant state $|\tau_{o}\rangle$ can be expressed by
\begin{equation}
\begin{aligned}
|\tau_{o}\rangle&={\mathcal O}|\tau_{init} \rangle \\
&=\frac{1}{\sqrt{M}}\sum_{j=1}^M|j\rangle \frac{1}{\sqrt{U}}\sum_{i=1}^U|i\rangle|v_{ji}\rangle|0\rangle.
\end{aligned}
\end{equation}
After that, an unitary operation
\begin{equation}
R_y(2{\rm sin}^{-1}v_{ji})=
\left[ \begin{array}{cc}
\sqrt{1-v^2_{ji}} & -v_{ji}\\
v_{ji} & \sqrt{1-v^2_{ji}}
\end{array}
\right ]
\end{equation}
is applied to the last quantum bit (qubit) of $|\tau_{o}\rangle$, so that the rotated state 
\begin{equation}
\begin{aligned}
|\tau_{r}\rangle&=R_y(2{\rm sin}^{-1}v_{ji}) |\tau_{o}\rangle \\ &=\frac{1}{\sqrt{M}}\sum_{j=1}^M|j\rangle \frac{1}{\sqrt{U}}\sum_{i=1}^U|i\rangle|v_{ji}\rangle(\sqrt{1-v_{ji}^2}|0\rangle+v_{ji}|1\rangle)
\end{aligned}
\end{equation}
can be obtained. Finally, the quantum state $|\tau\rangle$ can be obtained by removing the auxiliary qubit $|v_{ji}\rangle$ of $|\tau_{r}\rangle$ using Oracle ${\mathcal O}^{\dagger}$. Similarly, the quantum state $|\rho\rangle$ can be prepared in the same way with initial state $|\rho_{init}\rangle=\frac{1}{\sqrt{U}}\sum_{i=1}^U|i\rangle|0\rangle|0\rangle$. As a result, the information of feature vectors can be stored in the amplitude of quantum states $|\tau\rangle$ and $|\rho\rangle$.

\begin{figure}
\centering
\includegraphics[scale=0.8]{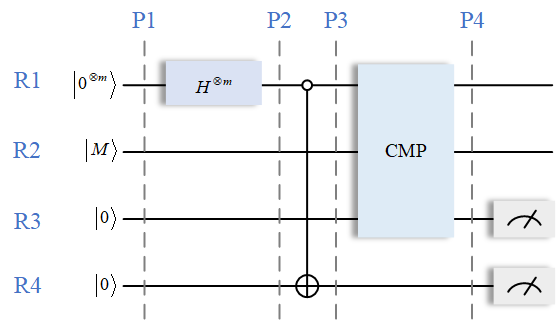}
\caption{The quantum circuit for preparing $\frac{1}{\sqrt{M}}\sum_{j=1}^M|j\rangle$. $|M\rangle$ is the computational basis state that stores $M$ as a binary string, and $m = \lceil {\rm log}_2(M + 1)\rceil$. $H$ is the Hadamard gate, CMP is a comparison operator.}
\label{fig:inintialstate}
\end{figure}

\subsection{Similarity calculating}\label{SC}

\begin{figure}
\centering
\includegraphics[scale=0.8]{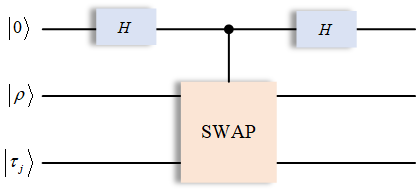}
\caption{The c-SWAP test quantum circuit.}
\label{fig:SWAP}
\end{figure}

For now, Bob possesses both the quantum state $|\tau\rangle$ that carries the information of known DMCSs and the quantum state $|\rho\rangle$ that carries the information of unknown DMCS. He first calculates the fidelity, which can be used as a measure of the similarity, between these quantum states and store it to the amplitude of qubit by performing a controlled-SWAP (c-SWAP) test shown in Fig.~\ref{fig:SWAP}. The quantum circuit of c-SWAP test is composed of two Hadamard gate and a SWAP operation which obeys ${\rm SWAP}|\rho\rangle|\tau_j\rangle =|\tau_j\rangle|\rho\rangle$, where 
\begin{equation}\label{equ:tauj}
\begin{aligned}
|\tau_j \rangle=\frac{1}{\sqrt{U}}\sum_{i=1}^U|i\rangle|1\rangle(\sqrt{1-v_{ji}^2}|0\rangle+v_{ji}|1\rangle).
\end{aligned}
\end{equation}
Therefore, the function of c-SWAP test is that the last two states will be swapped if the control qubit is $|1\rangle$ or they will not be swapped if the control qubit is $|0\rangle$. 

After passing the c-SWAP test, the input state can be transformed to
\begin{equation}\label{equ:Phy}
\begin{aligned}
|\gamma_j\rangle&=\mathrm{c\mbox{-}SWAP}|0\rangle|\rho\rangle|\tau_j\rangle \\
&=\frac{1}{2}|0\rangle(|\rho\rangle|\tau_j\rangle+|\tau_j\rangle|\rho\rangle)+\frac{1}{2}|1\rangle(|\rho\rangle|\tau_j\rangle-|\tau_j\rangle|\rho\rangle)\\
&=\sqrt{P_j(0)}|0\rangle+\sqrt{1-P_j(0)}|1\rangle, 
\end{aligned}
\end{equation}
where 
\begin{equation}\label{Pj0}
\begin{aligned}
P_j(0)=\frac{1+|\langle\rho|\tau_j\rangle|^2}{2}.
\end{aligned}
\end{equation}
See Appendix \ref{P0} for the detailed derivation of $P_j(0)$. It is easy to find that $P_j(0)$ is directly proportional to the fidelity $|\langle\rho|\tau_j\rangle|^2$, so that $P_j(0)$ can be directly used for measuring the similarity between $|\rho\rangle$ and $|\tau_j\rangle$. After $M$ times c-SWAP test, a superposition state whose amplitudes contain all similarities can be finally obtained as
\begin{equation}\label{equ:gamma}
\begin{aligned}
|\gamma\rangle = \frac{1}{\sqrt{M}}\sum_{j=1}^M|j\rangle(\sqrt{P_j(0)}|0\rangle+\sqrt{1-P_j(0)}|1\rangle).
\end{aligned}
\end{equation}

\begin{figure}
\centering
\includegraphics[scale=0.8]{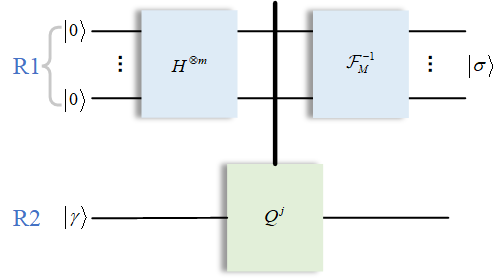}
\caption{The quantum circuit for amplitude estimation, which is used to obtain the estimate from the amplitude of a quantum state. $Q$ is an unitary operator which obeys $Q=-{\mathcal A}S_0{\mathcal A}^{\dagger}S_{\chi}$ and ${\mathcal F}^{-1}$ is quantum inverse Fourier transform. $Q^j$ denotes $j$ applications of operator $Q$.}
\label{fig:AE}
\end{figure}

In order to facilitate the processing of follow-up quantum search algorithm, amplitude estimation \cite{Brassard_2002} is further applied to this superposition state so that the amplitudes of $|\gamma\rangle$ can be stored as a qubit string. Figure \ref{fig:AE} shows the quantum circuit for implementing amplitude estimation, which includes two steps i.e., amplitude amplification and phase estimation. Specifically, amplitude amplification is implemented by an unitary operator $Q=-{\mathcal A}S_0{\mathcal A}^{\dagger}S_{\chi}$ where $\mathcal A$ performs $|0^{\otimes m}\rangle\rightarrow|\gamma\rangle$, $S_0$ obeys
\begin{equation}
S_0|x\rangle = \left\{
\begin{aligned}
|x\rangle , x\neq 0\\
-|x\rangle , x=0\\
\end{aligned}
\right
.,
\end{equation}
and $S_{\chi}$ obeys
\begin{equation}
S_{\chi}|x\rangle = \left\{
\begin{aligned}
|x\rangle , \chi(x)=0\\
-|x\rangle , \chi(x)=1\\
\end{aligned}
\right
.,
\end{equation}
while phase estimation is implemented by an inverse quantum Fourier transform (IQFT) ${\mathcal F}^{-1}$ which is defined as
\begin{equation}
\begin{aligned}
{\mathcal F}_M^{-1} |x\rangle = \frac{1}{\sqrt{M}}\sum_{y=0}^{M-1}e^{-\iota\frac{2\pi }{M}xy}|y\rangle,
\end{aligned}
\end{equation}
where $0\le x<M$, $\iota=\sqrt{-1}$.
After amplitude estimation, a quantum state 
\begin{equation}
\begin{aligned}
|\sigma\rangle=\frac{1}{\sqrt{M}}\sum_{j=1}^M|j\rangle \left|{\rm Sim}_j\right\rangle
\end{aligned}
\end{equation}
can be prepared, where ${\rm Sim}_j=\frac{M}{\pi}{\rm arcsin}(\sqrt{\widetilde{P_j}(0)})$ and $\widetilde{P_j}(0)$ is the estimate of $P_j(0)$. See Appendix \ref{AE} for the detailed derivation of quantum state $|\sigma\rangle$. For now, the similarities can be deemed to have been stored as a qubit string due to ${\rm Sim}_j$ is also directly proportional to the fidelity $|\langle\rho|\tau_j\rangle|^2$.

\subsection{Quantum $k$-maximal finding}\label{MF}

Bob now holds the quantum state $|\sigma\rangle$ whose qubit string contains the information of similarities between all known DMCSs and the unknown DMCS. He then finds $k$ nearest neighbors of the unknown DMCS by applying quantum $k$-maximal finding algorithm, which is detailed as follows.

\begin{figure}
\centering
{\includegraphics[scale=0.75]{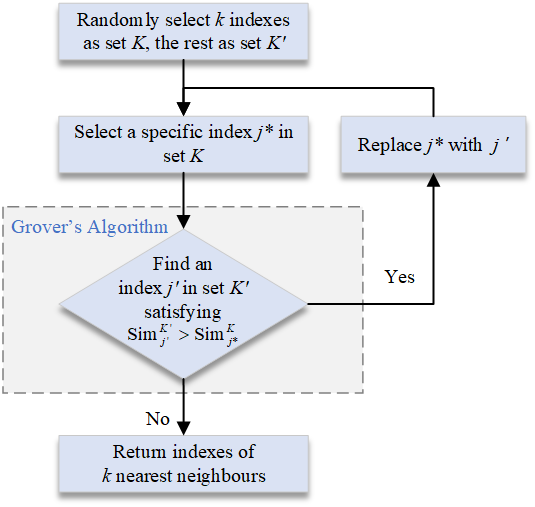}}
\caption{The flow chart of the quantum $k$-maximal finding algorithm. }
\label{fig:searching}
\end{figure}

\begin{figure*}
\centering
\includegraphics[scale=0.75]{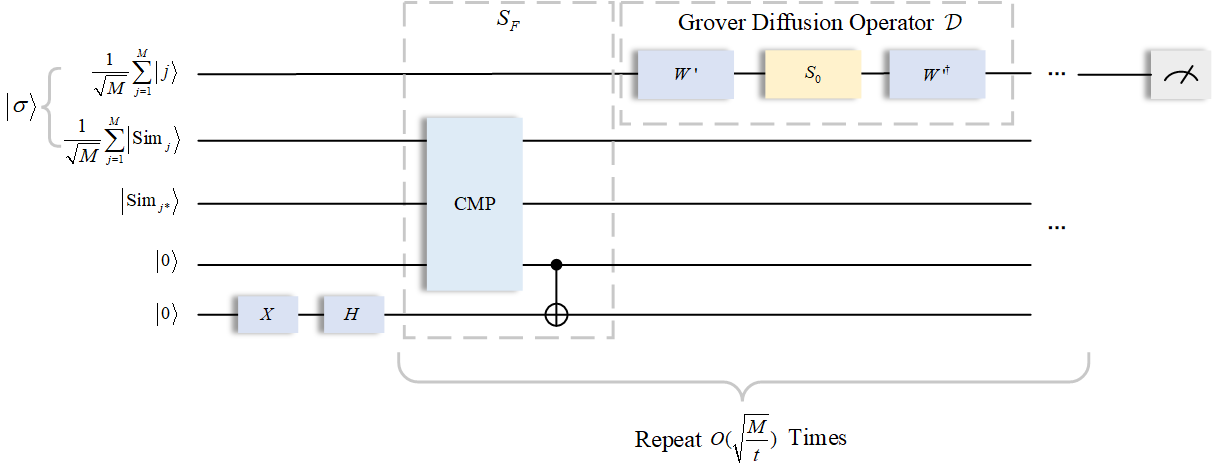}
\caption{Quantum circuit of Grover's algorithm. The Grover iteration is composed of an unitary transformation $S_F$ and Grover diffusion operator ${\mathcal D}$. $S_F$ is an unitary transformation to realize phase inversion of the target quantum state, and Grover diffusion operator ${\mathcal D}$ is an operator to realize inversion about average of the quantum state. It needs to repeat $O(\sqrt{\frac{M}{t}})$ times of Grover iterations to complete Grover's algorithm.}
\label{fig:Grover1}
\end{figure*}

As shown in Fig.~\ref{fig:searching}, Bob first randomly selects $k$ known DMCSs and records their corresponding indexes $j$ into an initial set $K$, the rest indexes $j'$ are recorded into a set $K'$. 
Then, the Grover's algorithm \cite{grover1996fast} is applied to $|\sigma\rangle$ to find an index $j'$ whose corresponding $|{\rm Sim}_{j'}^{K'}\rangle$ satisfying ${\rm Sim}_{j'}^{K'}>{\rm Sim}_{j^*}^K$, where $j^*\in K$.
In our case, the quantum circuit for implementing Grover's algorithm is shown in Fig.~\ref{fig:Grover1}, which includes a number of Grover iterations and a final measurement. Each iteration is composed of a unitary operator $S_F$ and a Grover diffusion operator ${\mathcal D}$. Specifically, $S_F$ denotes conditional phase shift transformation which obeys
\begin{equation}
S_F|j\rangle = \left\{
\begin{aligned}
-&|j\rangle , F(j)=1\\
&|j\rangle , {\rm otherwise}\\
\end{aligned}
\right
.,
\end{equation}
where the function $F$ is defined as
\begin{equation}
F(j) = \left\{
\begin{aligned}
&1 ,\;{\rm Sim}_{j}>{\rm Sim}_{j^*}^K\\
&0 ,\;{\rm otherwise}\\
\end{aligned}
\right
..
\end{equation}
The $S_F$ operator can be implemented by a CMP operator and a CNOT gate. After passing the CMP operation, the fourth input qubit $|0\rangle$ is transformed to
\begin{equation}
\begin{aligned}
\frac{1}{\sqrt{M}}\sum_{F(j)=1}|1\rangle+\frac{1}{\sqrt{M}}\sum_{F(j)\neq1}|0\rangle.
\end{aligned}
\end{equation}
Then, the CNOT gate is controlled by the above state so that the last input qubit $|0\rangle$ is transformed to
\begin{equation}
\begin{aligned}
|0\rangle&\stackrel{X}{\longrightarrow} |1\rangle\stackrel{H}{\longrightarrow} \frac{1}{\sqrt{2}}(|0\rangle-|1\rangle)\\
&\stackrel{CNOT}{\longrightarrow} \frac{1}{\sqrt{2M}}\sum_{j=1}^M(-1)^{F(j)}(|0\rangle-|1\rangle).
\end{aligned}
\end{equation}
Therefore, the resultant state 
\begin{equation}\label{equ:21}
\begin{aligned}
&\frac{1}{\sqrt{M}}\sum_{j=1}^M|j\rangle(-1)^{F(j)}\otimes\frac{1}{\sqrt{2}}(|0\rangle-|1\rangle)\\
=&\left(-\frac{1}{\sqrt{M}}\sum_{F(j)=1}|j\rangle+\frac{1}{\sqrt{M}}\sum_{F(j)\neq1}|j\rangle\right)\otimes\frac{1}{\sqrt{2}}(|0\rangle-|1\rangle)
\end{aligned}
\end{equation}
can be obtained after applying $S_F$ operator.
The Grover diffusion operator ${\mathcal D}=W'S_0W'^{\dagger}$ is subsequently applied to the first qubit of the resultant state,
so that its amplitude will be inversed about average \cite{PhysRevLett.79.325}. After the first Grover iteration, the state
\begin{equation}\label{equ:22}
\begin{aligned}
\frac{3M-4t}{M}\frac{1}{\sqrt{M}}\sum_{F(j)=1}|j\rangle+\frac{M-4t}{M}\frac{1}{\sqrt{M}}\sum_{F(j)\neq1}|j\rangle,
\end{aligned}
\end{equation}
where $t$ is the number of solutions of Grover's algorithm, can be obtained. From Eq.~(\ref{equ:21}) and Eq.~(\ref{equ:22}), we can easily find that the amplitude of $\sum_{F(j)=1}|j\rangle$ is amplified. After enough rounds of Grover iteration, the index $j'$ can be obtained by the final measurement with a probability approaching 1 (Details about this iteration is presented in Appendix \ref{Grover}). Bob then replaces $j^*$ with $j'$, and starts a new round of Grover's algorithm. Finally, the indexes of $k$ nearest neighbors can be obtained from the final set $K$.

\section{Performance analysis and discussion}

In this section, we first present the performance of the proposed Q$k$NN algorithm, followed by its complexity analysis. Subsequently, the security of the whole Q$k$NN-based CVQKD scheme is detailedly analyzed.

\subsection{Performance of Q$k$NN algorithm}\label{4.1}

\begin{table}[b]
\caption{Confusion matrix.}
\begin{ruledtabular}\label{tab:confusion}
\begin{tabular}{c c c}
& Predicted positive & Predicted negative \\
\hline
Actual positive & $\mathcal{N}_{TP}$ & $\mathcal{N}_{FN}$\\

Actual negative & $\mathcal{N}_{FP}$ & $\mathcal{N}_{TN}$
\end{tabular}
\end{ruledtabular}
\end{table}
As the proposed Q$k$NN can be deemed a quantum version of $k$NN algorithm, classical machine learning metrics can be used for evaluating its performance. In machine learning area, the confusion matrix is one of the most widely used concepts to analyze the performance of a certain classification algorithm. Table \ref{tab:confusion} shows the confusion matrix, in which the true positive ($\mathcal{N}_{TP}$) indicates the counts that actual positive data are predicted as positive, false positive ($\mathcal{N}_{FP}$) indicates the counts that actual negative data are predicted as positive, false negative ($\mathcal{N}_{FN}$) indicates the counts that actual positive data are predicted as negative, and true negative ($\mathcal{N}_{TN}$) indicates the counts that actual negative data are predicted as negative. 

\begin{figure}
\centering
\includegraphics[scale=0.27]{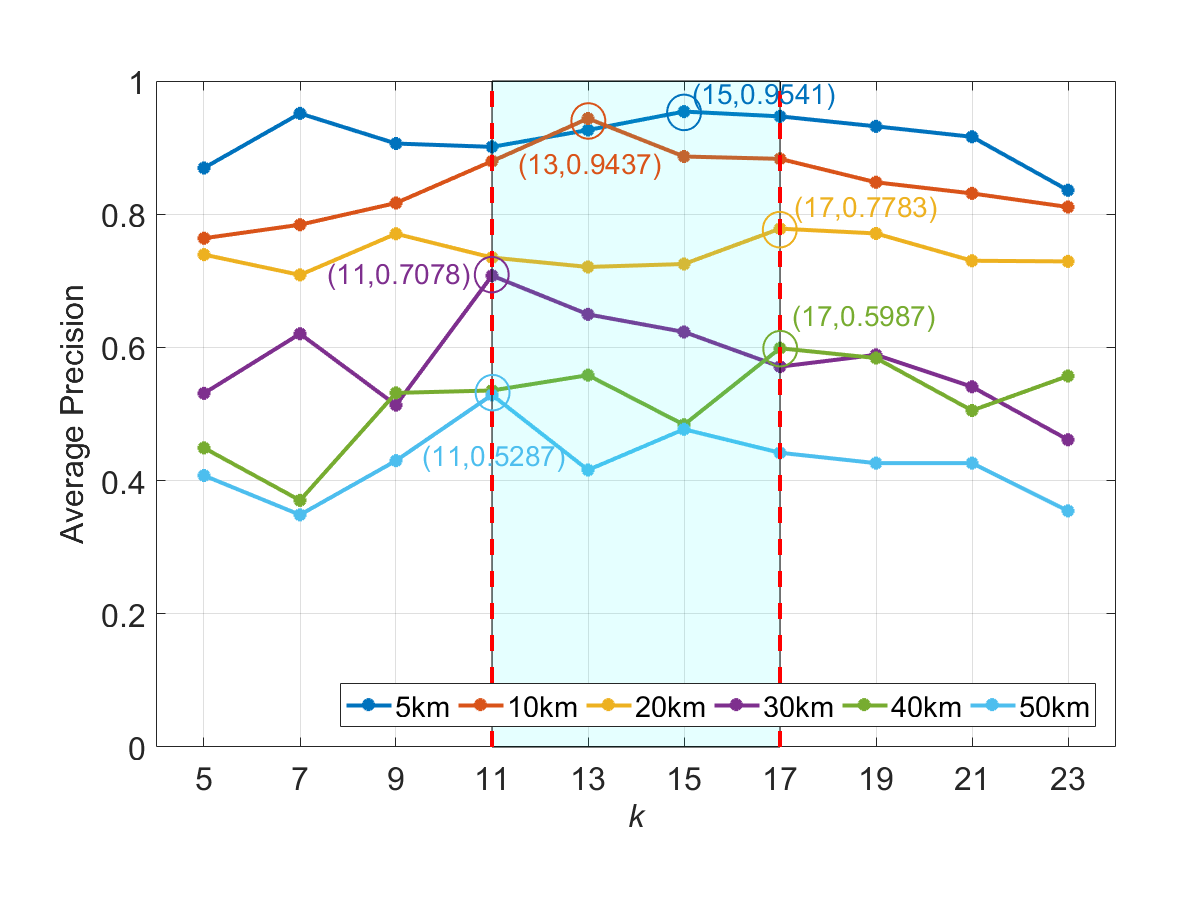}
\caption{The average precision $\bar{\mathcal{P}}_{\rm Prec}$ of Q$k$NN algorithm as a function of $k$. From top to bottom, the broken lines represent the distance of 5 km, 10 km, 20 km, 30 km, 40 km and 50 km, respectively.}
\label{fig:accuracy}
\end{figure}

With confusion matrix, a machine learning metric called {\bf Precision} ($\mathcal{P}_{\rm Prec}$) can be defined as 
\begin{equation}\label{equ:prec}
\begin{aligned}
\mathcal{P}_{\rm Prec} = \frac{\mathcal{N}_{TP}}{\mathcal{N}_{TP}+\mathcal{N}_{FP}},
\end{aligned}
\end{equation}
which indicates the proportion of actual positive data in all predicted positive data. This metric is important to our case as raw key is generated from the correct labels in all the predicted labels. Since there are eight labels existed in phase space, we further calculate the precision of each label $\mathcal{P}_{{\rm Prec}_i}$ and average them to obtain a metric {\bf Average Precision} $\bar{\mathcal{P}}_{\rm Prec}$, i.e.
\begin{equation}
\begin{aligned}
\bar{\mathcal{P}}_{\rm Prec} = \frac{1}{8}\sum_{i=1}^8 \mathcal{P}_{{\rm Prec}_i}.
\end{aligned}
\end{equation}
Obviously, the higher the value of $\bar{\mathcal{P}}_{\rm Prec}$, the more correlated the raw key.
Figure~\ref{fig:accuracy} shows the average precision $\bar{\mathcal{P}}_{\rm Prec}$ of the proposed Q$k$NN algorithm with different hyper-parameter $k$. By and large, it can be found that the value of $\bar{\mathcal{P}}_{\rm Prec}$ decreases with transmission distance increases, which indicates that the transmission distance is a crucial factor that impacts the performance of Q$k$NN algorithm. This is in line with our expectation, as the increased transmission distance will lead to more channel losses, resulting in extremely lower SNR of the received DMCSs. In addition, we find that $\bar{\mathcal{P}}_{\rm Prec}$ fluctuates as the hyper-parameter $k$ varies, and this situation exists in all distances. To select a proper $k$, we mark the peak value of each line out with a circle. It is observed that these peak values occur when $k$ is set from 11 to 17, e.g. $\bar{\mathcal{P}}_{\rm Prec}=0.9541$ at transmission distance of $5$ km when $k=15$, which suggests that the value of $k$ should not be set too small or too large. This is because a small value of $k$ may lead to overfitting problem and a large value of $k$ may result in underfitting problem, while $\bar{\mathcal{P}}_{\rm Prec}$ will be decreased by either of the two problems \cite{hastie2001elements}. 

\begin{figure}
\centering
\includegraphics[scale=0.42]{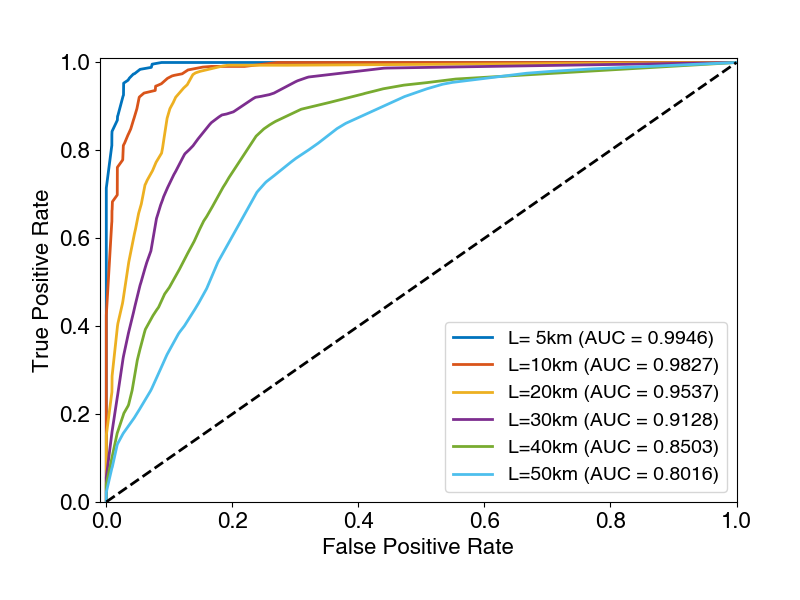}
\caption{The macro-average ROC curves of Q$k$NN with hyper-parameter $k=15$. Dashed line denotes the performance of random guess.}
\label{fig:ROC}
\end{figure}

For now, we have investigated the performance of Q$k$NN in terms of precision. To comprehensively evaluate a classifier, however, a sole metric is inadequate. Therefore, an overall metric called macro-average {\bf Receiver Operating Characteristic} (ROC) curve \cite{FAWCETT2006861}, which describes the average {\bf True Positive Rate} ($\mathcal{P}_{\rm TPR}$) of a certain multi-class classifier as a function of its average {\bf False Positive Rate} ($\mathcal{P}_{\rm FPR}$), is introduced. For each class, $\mathcal{P}_{\rm TPR}$ indicates the proportion of the data that is correctly predicted as positive in all actual positive data, and $\mathcal{P}_{\rm FPR}$ indicates the proportion of the data that is incorrectly predicted as positive in all actual negative data. Their formulas are given by
\begin{equation}\label{equ:tpr}
\begin{aligned}
\mathcal{P}_{\rm TPR} = \frac{\mathcal{N}_{TP}}{\mathcal{N}_{TP}+\mathcal{N}_{FN}},
\end{aligned}
\end{equation}
\begin{equation}\label{equ:fpr}
\begin{aligned}
\mathcal{P}_{\rm FPR} = \frac{\mathcal{N}_{FP}}{\mathcal{N}_{FP}+\mathcal{N}_{TN}}.
\end{aligned}
\end{equation}
Figure \ref{fig:ROC} shows the macro-average ROC curves of the proposed Q$k$NN with hyper-parameter $k=15$. The dashed line is the result of random guess, which illustrates that there is no performance improvement without using any classiﬁcation algorithm. With macro-average ROC curve, one can explicitly tell the quality of the multi-class classiﬁer: the curve more close to point $(0, 1)$, the performance better. Obviously, the proposed Q$k$NN can dramatically improve the performance of predicting DMCSs. Moreover, it can be observed that these curves are away from the best point $(0, 1)$ with the transmission distance increases, this trend is identical with our previous analysis, which further illustrates that channel loss is important for the performance of Q$k$NN algorithm. Besides, we further calculate the {\bf area under curve} (AUC) \cite{BRADLEY19971145} for each distance and thus obtain AUC value $\Lambda_{\rm Q}$, which is a probability value range from 0 to 1. As a numerical value, $\Lambda_{\rm Q}$ can be directly used for quantitatively evaluating classiﬁer’s quality. It can be easily found that our proposed Q$k$NN algorithm achieves extremely high overall performance ($\Lambda_{\rm Q}=0.9946$) with transmission distance is $5$ km. It is worth noting that the AUC value of random guess is 0.5, while it gets 0.8016 with the proposed Q$k$NN even the transmission distance is increased to $50$ km. This result demonstrates that the effectiveness of Q$k$NN in CVQKD system, therefore, the AUC value $\Lambda_{\rm Q}$ can be used to describe the efficiency of quantum classifier in the following security analysis of the proposed Q$k$NN-based CVQKD system.

\subsection{Complexity analysis for Q$k$NN algorithm}\label{4.2}

\begin{figure*}
\centering
\subfigure[\label{fig:complexity_bar}]{\includegraphics[scale=0.27]{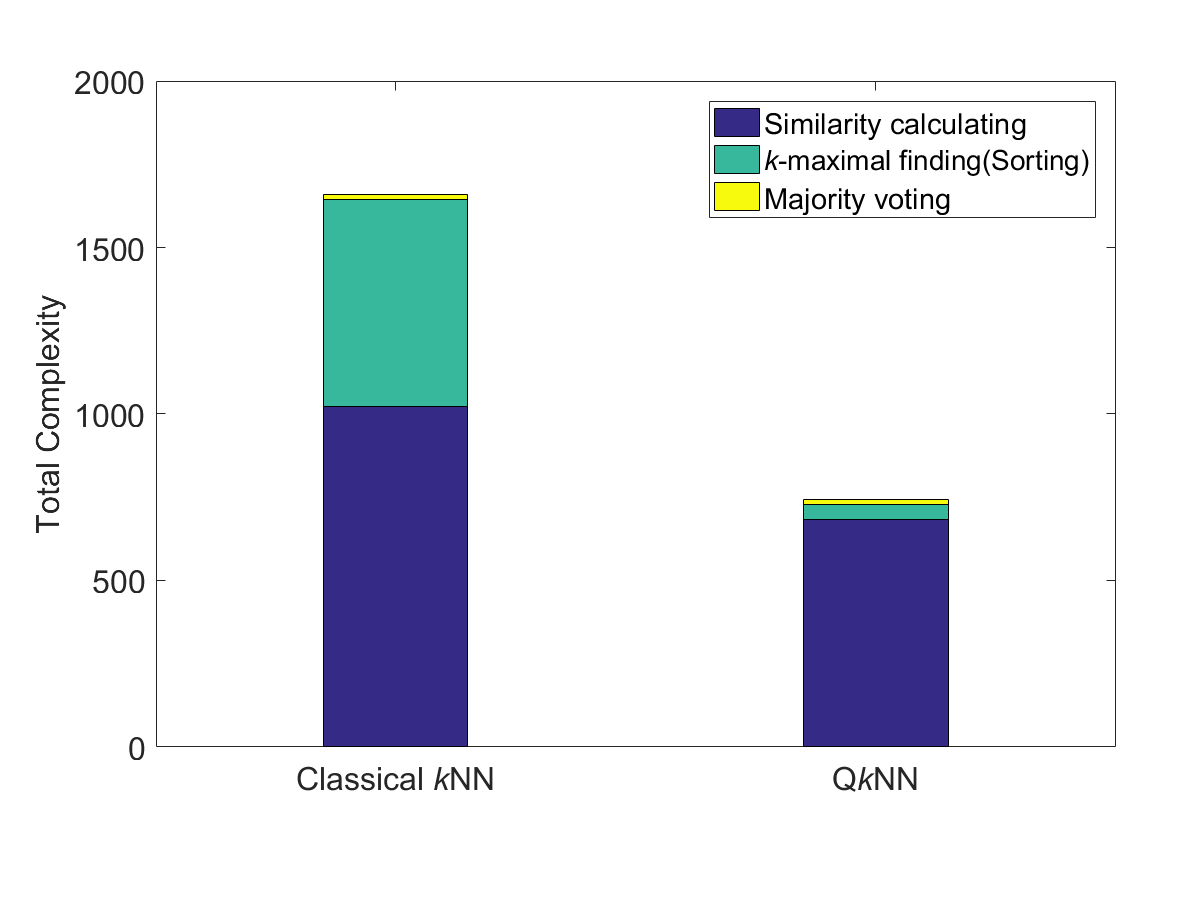}}
\subfigure[\label{fig:complexity_U}]{\includegraphics[scale=0.27]{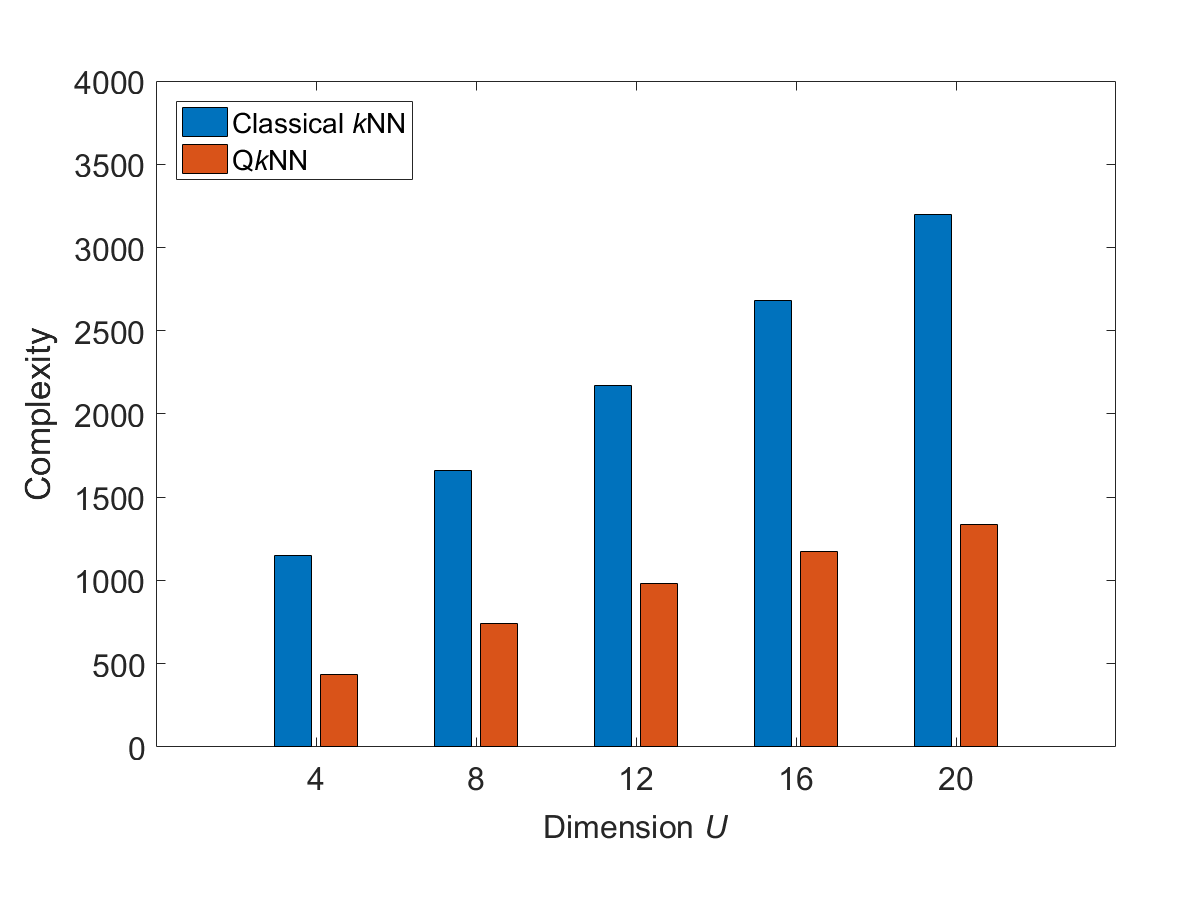}}
\\\vspace{-3mm}
\subfigure[\label{fig:complexity_M}]{\includegraphics[scale=0.27]{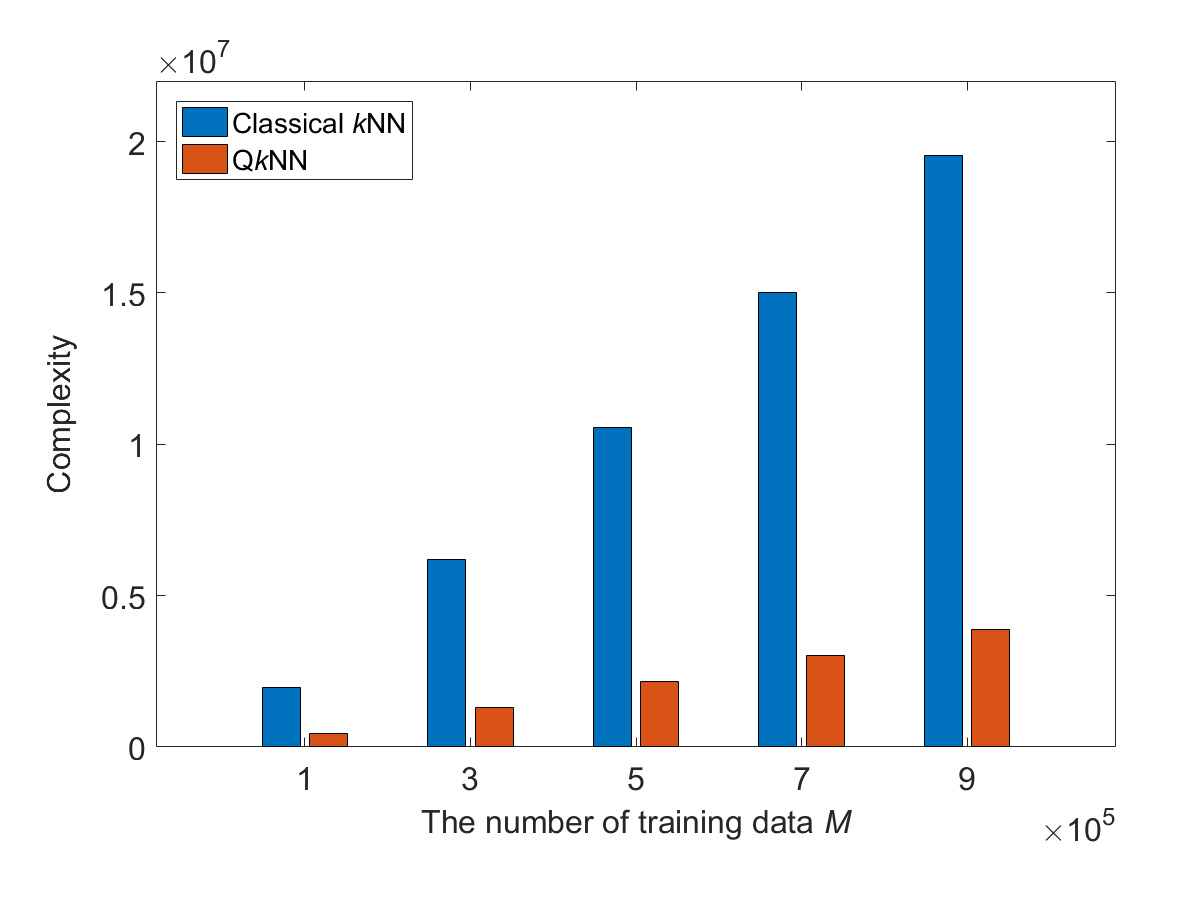}}
\subfigure[\label{fig:complexity_k}]{\includegraphics[scale=0.27]{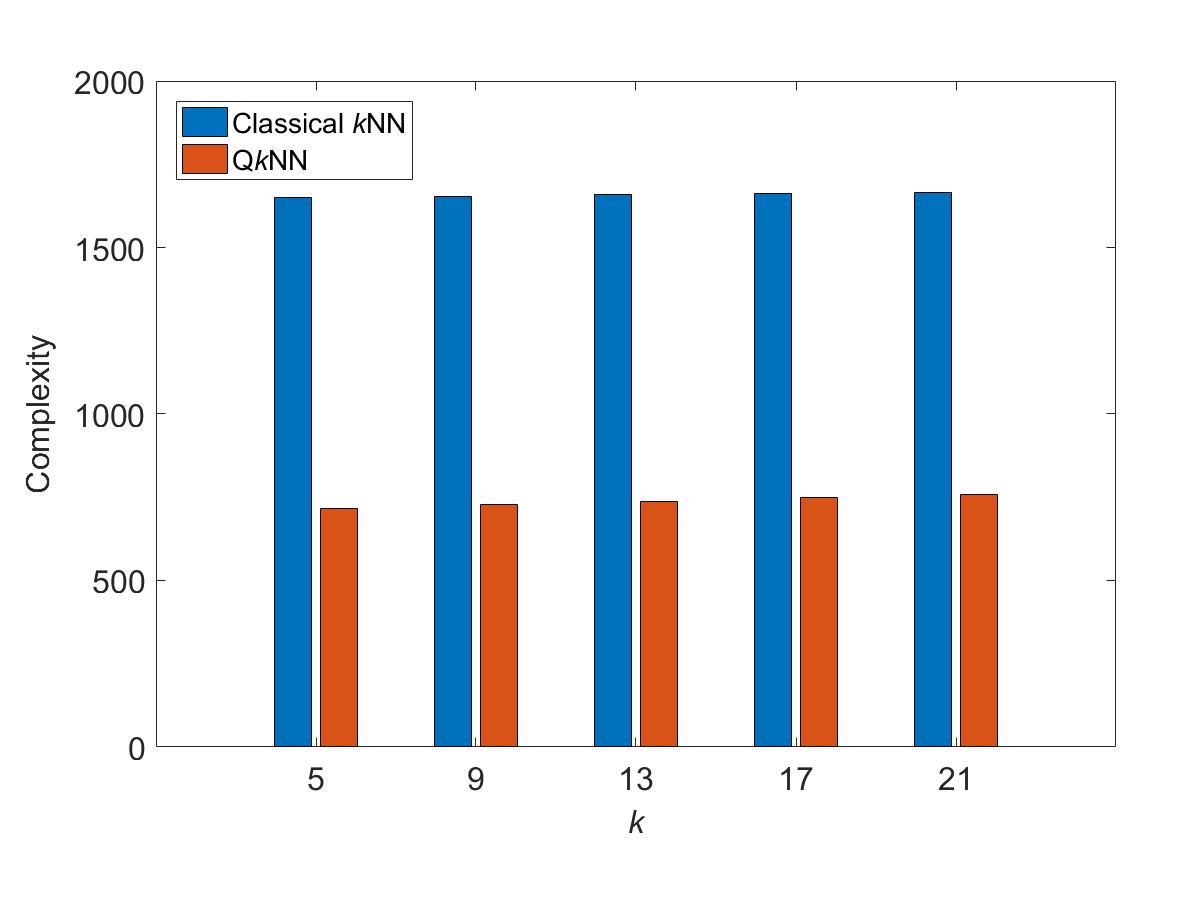}}
\caption{Complexity comparisons. (a) Comparison of total complexity. Parameters are set to be as follows. $U=8$, $R=131$ $(\delta=0.1)$, $M=128$ and $k=15$; (b) Complexity comparison as a function of dimension $U$; (c) Complexity comparison as a function of $M$; (d) Complexity comparison as a function of hyper-parameter $k$. }
\label{fig:complexity}
\end{figure*}

Before presenting the complexity analysis of Q$k$NN, let us briefly retrospect the complexity of classical $k$NN first. Assuming there are $M$ training data points $V_j=\{{\bf v_1},{\bf v_2},...,{\bf v_M}\}$ and each training data point is represented as an $U$ dimensional feature vector ${\bf v_j}=(v_{j1},v_{j2},...,v_{jU})$, the classical $k$NN algorithm first needs to calculate all similarities between the unlabeled data point ${\bf v_0}$ and each labeled data point ${\bf v_j}$, so that the complexity of this step is $O(UM)$. Then, the $M$ similarities need to be sorted with a certain sorting algorithm such as quick sort, merge sort, heap sort, etc. In general, the complexity of above sorting algorithms is no less than $O(M{\rm log}_2M)$ \cite{cormen2022introduction}. Finally, the majority voting requires $O(k)$ times to count the number of labels and assigns the label to ${\bf v_0}$. As a result, the total complexity of classical $k$NN can be expressed by
\begin{equation}\label{equ:com_knn}
O(UM+M{\rm log}_2M+k).
\end{equation}

In what follows, let us discuss the complexity of the proposed Q$k$NN algorithm in detail. As shown in Fig.~\ref{fig:QkNN_QKD}, the whole Q$k$NN is composed of four parts, i.e., preparation of quantum states, similarity calculating, quantum $k$-maximal finding and majority voting.
For preparation of quantum states, as we detailed in Sec.~\ref{sec3a}, both quantum states $|\tau\rangle$ and $|\rho\rangle$ are prepared by 3 Oracles (i.e. $\mathcal{O}$, $R_y$ and $\mathcal{O}^{\dagger}$ ), so that the complexity of this part is $O(6)$.
For similarity calculating, the two quantum states $|\tau\rangle$ and $|\rho\rangle$ are first passed through the c-SWAP test quantum circuit to obtain the quantum state $|\gamma_j\rangle$ that contains the similarity $P_j(0)$. 
Note that the complexity of calculating a $P_j(0)$ is $O({\rm log}^2_2U)$ \cite{hai2022new}. To obtain the superposition state $|\gamma\rangle$ whose amplitudes contain all similarities, the c-SWAP test needs to be performed $M$ times, the complexity is thereby increased to $O(M{\rm log}^2_2U)$. Subsequently, the similarities need to be stored as a qubit string by amplitude estimation. Specifically, the $Q$ operator is repeatedly performed to estimate the amplitude $a$ (see details in Appendix \ref{AE}), and the error probability for estimating $a$ satisfies \cite{Brassard_2002}
\begin{equation}
\begin{aligned}
|a-\widetilde{a}|\leqslant \frac{\pi}{R}+\frac{\pi^2}{R^2},
\end{aligned}
\end{equation}
where $\widetilde{a}$ is the estimation of $a$, and $R$ is the iteration times of operator $Q$. Obviously, $R$ needs to satisfy the following inequation, i.e.
\begin{equation}
R\geqslant\frac{\pi(\pi+1)}{\delta},
\end{equation}
if the error probability $|a-\widetilde{a}|\leqslant \delta$. That is to say, the $Q$ operator needs to be performed at least $R$ times ($O(R)$) to ensure that the error probability is less than or equal to $\delta$. Therefore, the total complexity of similarity calculating is $O(M{\rm log}^2_2U+R)$.
For quantum $k$-maximal finding, let ${\mathcal T}$ be a set whose elements do not belong to set $K$ but are more similar to ${\bf v_0}$ than some points in set $K$, and the number of elements of set ${\mathcal T}$ is $t$. Obviously, to find out $k$ maximal values, the Grover's algorithm and replacement need to be repeatedly performed until set ${\mathcal T}$ is empty, i.e., $t=0$. Reference \cite{doi:10.1137/050644719} shows that $t$ can be reduced to $\frac{t}{2}$ by performing $O(k)$ iterations of Grover's algorithm and replacement when $t>2k$, i.e., the Oracle needs to preform $O(k\sqrt{\frac{M}{t}})$ times. Once $t$ is reduced to $t\le2k$, $O(\sqrt{\frac{M}{t}})$ times Oracles in each round of Grover's algorithm are required to ensure $t=0$. Therefore, 
\begin{equation}\label{equ_MF}
\left\{
\begin{aligned}
k(\sqrt{\frac{M}{2k}}+\sqrt{\frac{M}{4k}}+\sqrt{\frac{M}{8k}}+...) ,t>2k\\
\sum_{i=1}^{2k}\sqrt{\frac{M}{i}} , t\le2k
\end{aligned}
\right
.
\end{equation}
times Oracles are required for reducing $t$ to 0. From Eq.~(\ref{equ_MF}), we can easily find that the complexity of quantum $k$-maximal finding is $O(\sqrt{kM})$. 
Till now, we have presented the complexity analysis of the former three parts, note that the last part of Q$k$NN, namely the majority voting, is similar to that of classical $k$NN, so that its complexity remains $O(k)$. Therefore, the total complexity of our proposed Q$k$NN is
\begin{equation}\label{equ:com_qknn}
\begin{aligned}
O(M{\rm log}^2_2U+R+\sqrt{kM}+k).
\end{aligned}
\end{equation}

Figure \ref{fig:complexity} shows complexity comparisons between the proposed Q$k$NN and classical $k$NN. As can be seen in Fig.~\ref{fig:complexity}(a), the total complexity of the proposed Q$k$NN is much less than that of classical $k$NN, it illustrates that the proposed Q$k$NN algorithm could offer a significant speedup over classical $k$NN algorithm. To figure out how the acceleration happens, we further mark each part with different colors and the result shows that the complexities of similarity calculating and $k$-maximal finding (corresponding to sorting in classical $k$NN algorithm) are dramatically decreased by the proposed Q$k$NN. It illustrates that the quantum parts of Q$k$NN are crucial for speedup. In addition, we find that the complexities for both Q$k$NN and classical $k$NN algorithms are affected by several parameters, i.e., the dimension $U$, the number of training data $M$ and the number of nearest neighbors $k$. We therefore plot Fig.~\ref{fig:complexity}(b), Fig.~\ref{fig:complexity}(c) and Fig.~\ref{fig:complexity}(d) to investigate the respective influence of each parameter on complexity. As shown in Fig.~\ref{fig:complexity}(b) and Fig.~\ref{fig:complexity}(c), the complexity of classical $k$NN rises quite sharply with the increase of $U$ or $M$, while the complexity of Q$k$NN rises very slowly. It illustrates that the proposed Q$k$NN algorithm is of clear superiority in addressing high dimensional or large-size classification issues. Figure \ref{fig:complexity}(d) shows that although the complexity of classical $k$NN is apparently larger than that of Q$k$NN, the complexities of both algorithms are not sensitive to the hyper-parameter $k$, which illustrates that $k$ is not a crucial parameter that heavily affects the complexities of both classical $k$NN and Q$k$NN algorithms. It is worthy noting that, to explicitly show the respective trends, the scale of labeled data point $M$ is set to 128 in Fig.~\ref{fig:complexity}(a), Fig.~\ref{fig:complexity}(b) and Fig.~\ref{fig:complexity}(d). Actually, $M$ is usually far more than $10^5$ for a realistic CVQKD system \cite{Leverrier:2015he}. In such a practical scenario, the complexity gap between Q$k$NN and classical $k$NN will be extremely large.

\subsection{Security analysis of Q$k$NN-based CVQKD}\label{4.3}

\begin{figure}
\centering
\includegraphics[scale=0.27]{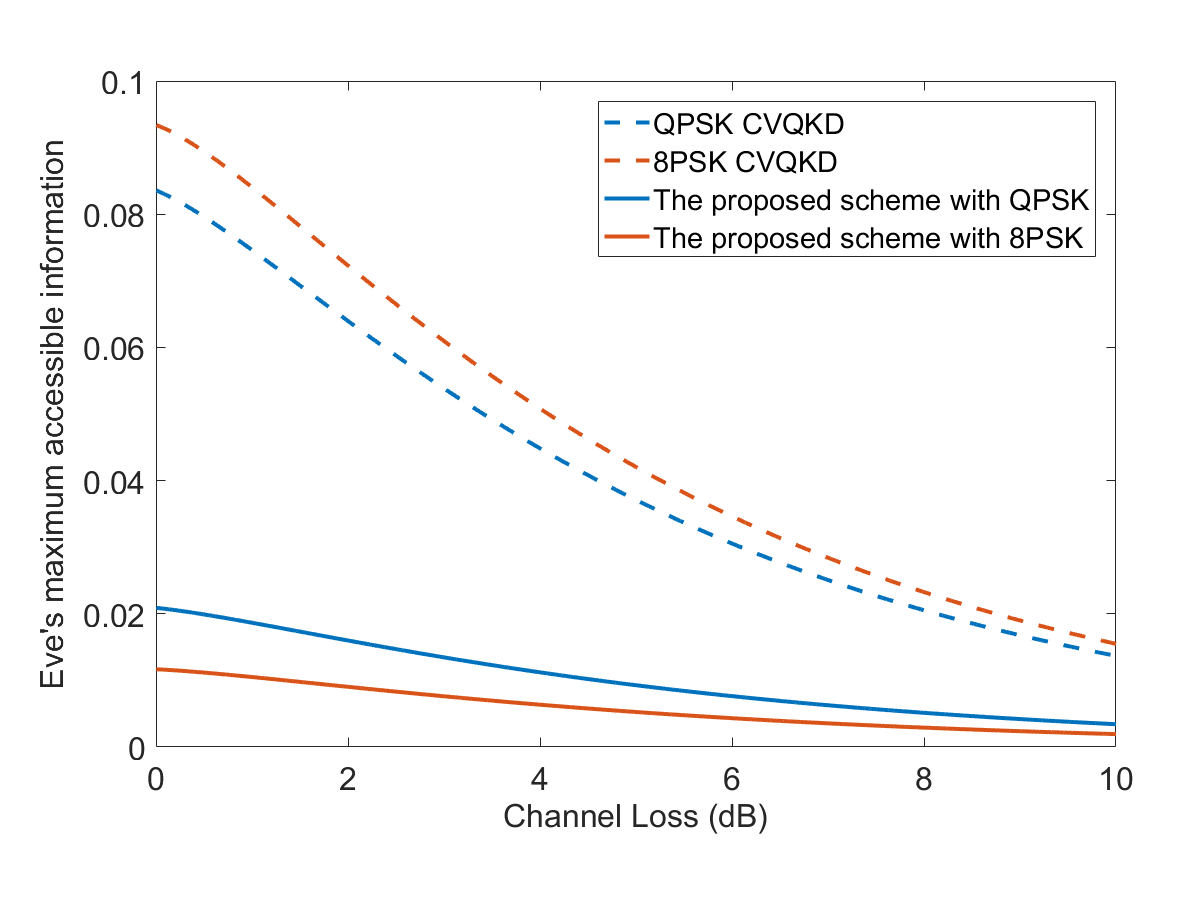}
\caption{Eve’s maximum accessible information as a function of channel loss (0.2 dB/km). Solid lines denote the proposed scheme with different QPSK and 8PSK, respectively. Dashed lines denote the conventional DM CVQKD with QPSK and 8PSK, respectively. Modulation variance $V_m$ are set to 0.33 for QPSK and 0.38 for 8PSK.}
\label{fig:Sbe}
\end{figure}

Till now, we have demonstrated the performance of Q$k$NN-based CVQKD in terms of machine learning metrics and have analyzed the complexity of Q$k$NN algorithm, both results have shown the advantages of our scheme. In what follows, we present the theoretical security proof for Q$k$NN-based CVQKD with semi-definite program (SDP) method \cite{Denys2021explicitasymptotic}, detailed calculations can be found in Appendix \ref{DMCVQKD}. As known, the asymptotic secret key rate of the conventional DM CVQKD with reverse reconciliation is given by \cite{10.1103/physreva.102.032604}
\begin{equation}
\begin{aligned}
K_{\rm asym} = \beta I_{\rm AB}-\chi_{\rm BE},\label{key}
\end{aligned}
\end{equation}
where $\beta$ is the reconciliation efficiency, $I_{\rm AB}$ is the Shannon mutual information between Alice and Bob, and $\chi_{\rm BE}$ is the Holevo bound of the mutual information between Eve and Bob. However, Eq.~(\ref{key}) does not consider the influence of the introduction of Q$k$NN classifier for both legitimate users (Alice and Bob) and the eavesdropper (Eve), it, therefore, has to be amended to suitable for evaluating quantum machine learning-based CVQKD. Due to the data processing is quite different, Eq.~(\ref{key}) can be rewritten as the following form 
\begin{equation}
\begin{aligned}
K_{\rm asym}^{\rm Q} = \beta \Lambda_{\rm Q} I_{\rm AB}-p(y_i)\chi_{\rm BE}.\label{keys}
\end{aligned}
\end{equation}
The difference between Eq.~(\ref{key}) and Eq.~(\ref{keys}) lies in two parts. First, the AUC value $\Lambda_{\rm Q}$ has to be considered as it describes the efficiency of quantum classifier. Higher AUC value implies higher correlation of raw key between Alice and Bob. Second, term $\chi_{\rm BE}$, which represents the Holevo quantity for Eve’s maximum accessible information, is reduced to $p(y_i)\chi_{\rm BE}$, where $p(y_i)=1/N$ is the probability of discrete uniform distribution when variable $Y = y_i(i = 1, 2, . . . , N)$. This is because Eve is no longer able to acquire as much information as before, due to some relationships are no longer fixed and public. To be specific, in conventional DM CVQKD, the relationship between each DMCS and its binary presentation is ﬁxed and public. For instance, the key bits of state $|\alpha_0\rangle$ in QPSK CVQKD is always (0, 0) (see Fig.~1 in Ref.~\cite{10.1103/physrevx.9.021059}), so that Eve can precisely recover the correct key bits (0, 0) when she successfully intercepts the coherent state $|\alpha_0\rangle$. Similarly for 8PSK CVQKD,  Eve can precisely recover the correct key bits (0, 0, 0) when she successfully intercepts the coherent state $|\alpha_0\rangle$. However, due to the special-designed process of Q$k$NN-based CVQKD, the above relationship is no longer ﬁxed and public. Although each DMCS is assigned to a fixed label, such as the label of $|\alpha_0\rangle$ is $L_1$ and the labels of $|\alpha_7\rangle$ is $L_8$ shown in Fig.~\ref{fig:label}, the binary presentation for each label can be randomly assigned by Alice. Since the initialization part is secure, Bob will learn the assignment by the transmitted DMCSs at the end of initialization. While Eve who does not participate the initialization part is completely unaware of the assignment shared by Alice and Bob. She thus can only guess the correct label with the success probability for $N$-PSK is $1/N$. As can be seen in Fig.~\ref{fig:Sbe}, in our proposed scheme, Eve’s maximum accessible information is apparently reduced when compared with conventional DM CVQKD, and it can be further decreased by using higher dimensional PSK modulation. This is opposite to the trend in conventional DM CVQKD in which Eve’s maximum accessible information will increase with higher dimensional PSK modulation. It illustrates that the proposed scheme can efficiently prevent eavesdropper from obtaining more useful information, thereby contributing to the increase of final secret key.

\begin{figure}
\centering
\includegraphics[scale=0.33]{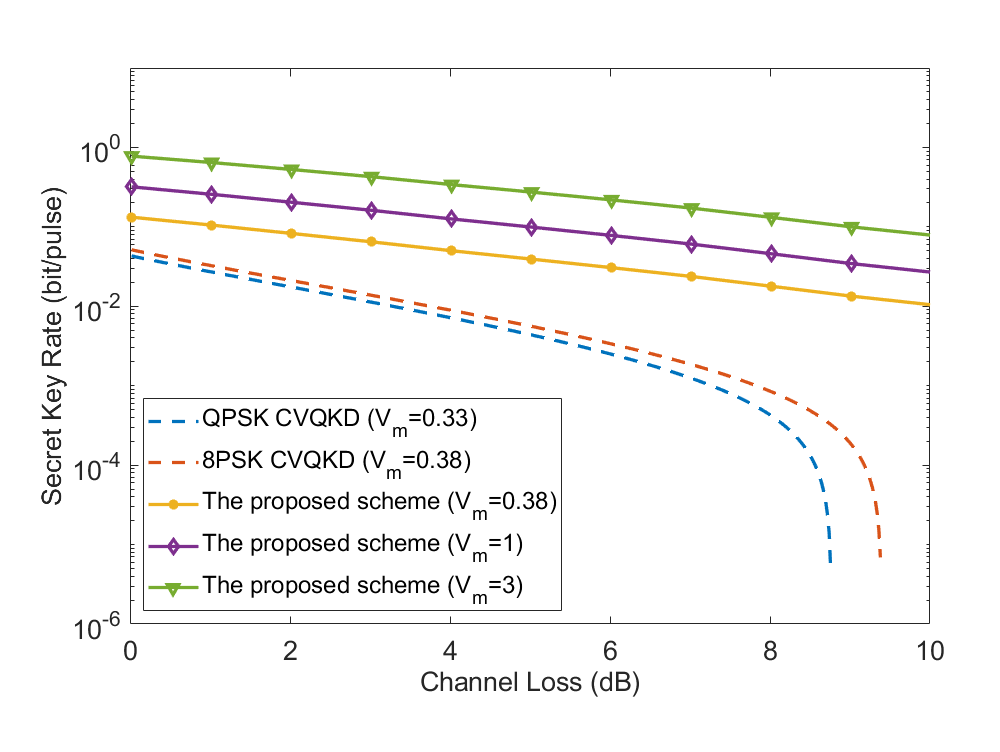}
\caption{Asymptotic secret key rate as a function of channel loss (0.2 dB/km). Solid lines denote the proposed Q$k$NN-based CVQKD with different modulation variances, blue dashed line denotes QPSK CVQKD, and red dashed line denotes 8PSK CVQKD. The parameters are set to be as follows. Detector efficiency $\eta = 0.6$, electronic noise $v_{el} = 0.05$, excess noise $\varepsilon = 0.01$, reconciliation efficiency $\beta = 0.98$ and $k=15$. }
\label{fig:SKR_cl}
\end{figure}
\begin{figure}
\centering
\includegraphics[scale=0.27]{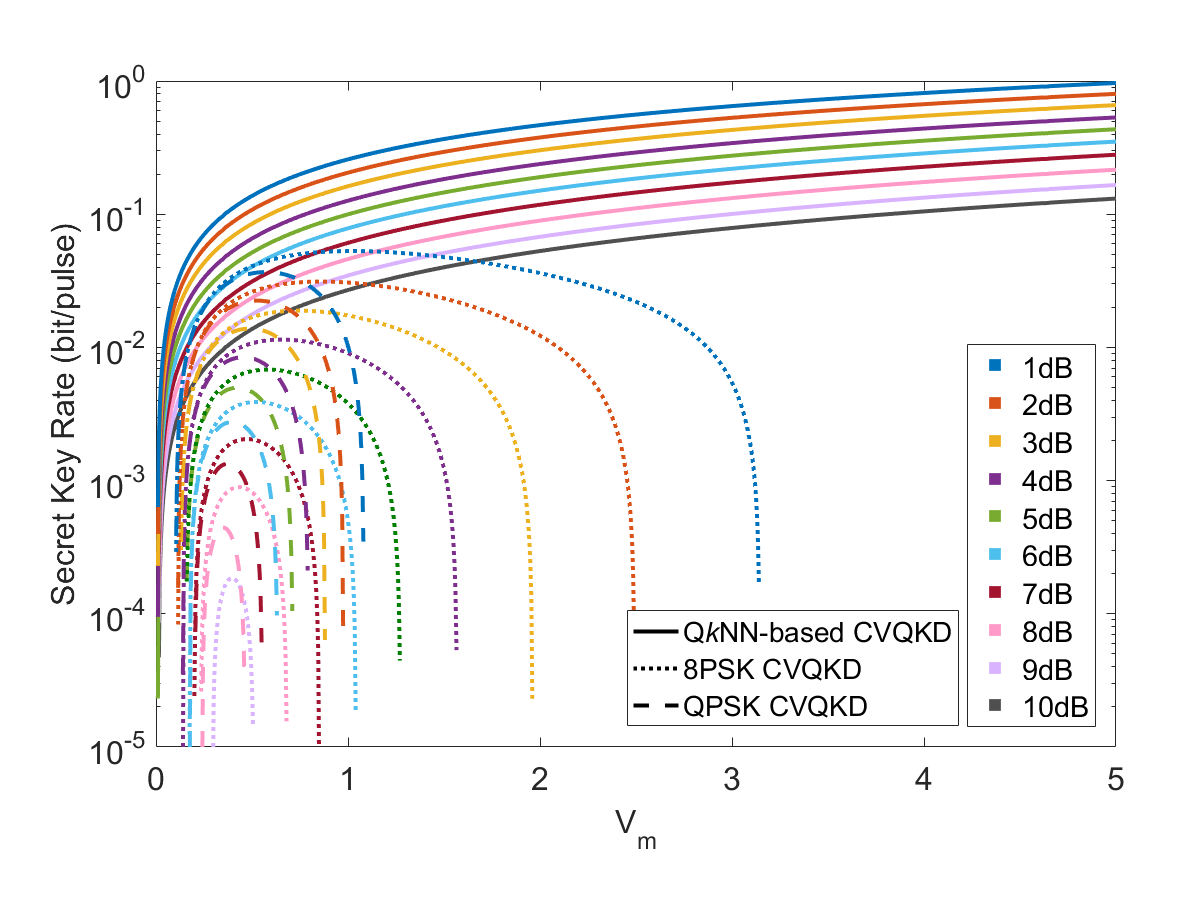}
\caption{Asymptotic secret key rate as a function of modulation variance $V_m$. Solid lines denote the Q$k$NN-based CVQKD, dotted lines denote 8PSK CVQKD, and dashed lines denote QPSK CVQKD. Different channel losses are marked with different colors.}
\label{fig:Vm_KeyRate}
\end{figure}
Figure \ref{fig:SKR_cl} shows the performance comparison between Q$k$NN-based CVQKD and two conventional DM CVQKD protocols in asymptotic limit. The results demonstrate that the secret key rate of Q$k$NN-based CVQKD outperforms other DM CVQKD protocols at all channel losses. In addition, we find that the secret key rate of the proposed scheme can be further increased with the risen modulation variance $V_m$. This is inconsistent with conventional DM CVQKD whose security has to be guaranteed by small modulation variance \cite{10.1103/physrevx.9.021059,10.1103/physreva.103.032410}. To investigate what caused this inconsistency, we plot Fig.~\ref{fig:Vm_KeyRate}, which shows the asymptotic secret key rates of above-mentioned schemes as a function of modulation variance. It can be easily found that the curves of both QPSK CVQKD and 8PSK CVQKD are arched and are located in certain ranges of small modulation variance, and the ranges are getting narrower with channel loss increases. Meanwhile, curves of Q$k$NN-based CVQKD are keep rising with the increase of modulation variance, it illustrates that small modulation variance is no longer needed for guaranteeing the security, so that the secret key rate of our proposed scheme can be further increased by setting proper larger modulation variance.

\section{Conclusion}

In this work, we have proposed a high-rate continuous-variable quantum key distribution scheme based on quantum machine learning, called Q$k$NN-based CVQKD. The proposed scheme divides the whole process of conventional DM CVQKD protocol into three parts, i.e., initialization, prediction and data-postprocessing. The initialization part is used for training and estimating quantum classifier, the prediction part is used for generating highly correlated raw keys, and
the data-postprocessing part generates the final secret key string shared by Alice and Bob. To this end, a specialized Q$k$NN algorithm was elegantly designed as a quantum classifier for distinguishing the incoming DMCSs. We then introduced several related machine learning-based metrics to estimate the performance of Q$k$NN, and compared its complexity with classical $k$NN algorithm. The asymptotic security proof of Q$k$NN-based CVQKD was finally presented with SDP method.

We have comprehensively analyzed the performance of Q$k$NN-based CVQKD, the results indicate that our proposed scheme is suitable for the high-speed metropolitan secure communication due to its advantages of high-rate and low-complexity. Besides, it is worthy noticing that the quantum classiﬁer is not limited to the proposed Q$k$NN, any other well-behaved quantum classifier can be used for improving CVQKD with the proposed processing framework.

In summary, Q$k$NN-based CVQKD is not only an improvement of CVQKD protocol, but also provides a novel thought for introducing various (quantum) machine learning-based methodologies to CVQKD ﬁeld.

\begin{acknowledgments}
 This work was supported by the National Natural Science Foundation of China (Grant No. 62101180), Hunan Provincial Natural Science Foundation of China (Grant No. 2022JJ30163), the Open Research Fund Program of the State Key Laboratory of High Performance Computing, National University of Defense Technology (Grant No. 202101-25).
\end{acknowledgments}

\begin{appendix}

\section{Derivation of the initial state}\label{SP}

To clearly describe the preparation of initial state, the quantum circuit shown in Fig.~\ref{fig:inintialstate} is divided into four phases P1-P4. Thereinto, $H$ is Hadamard gate which is defined by
\begin{equation}
H=\left[ \begin{array}{cc}
\frac{1}{\sqrt{2}} & \frac{1}{\sqrt{2}}\\
\frac{1}{\sqrt{2}} & -\frac{1}{\sqrt{2}}
\end{array}
\right ],
\end{equation}
we therefore have $H(|0\rangle)=\frac{1}{\sqrt{2}}(|0\rangle+|1\rangle)$ and $H(|1\rangle)=\frac{1}{\sqrt{2}}(|0\rangle-|1\rangle)$. CMP is an unitary operation which obeys
\begin{equation}\label{equ:CMP}
{\rm CMP}|i\rangle|M\rangle|0\rangle = \left\{
\begin{aligned}
|i\rangle|M\rangle|0\rangle \quad i\leqslant M\\
|i\rangle|M\rangle|1\rangle \quad i>M\\
\end{aligned}
\right
.,
\end{equation}
where $|M\rangle$ is the computational basis state that stores $M$ as a binary string.

\begin{figure}
\centering
\includegraphics[scale=0.9]{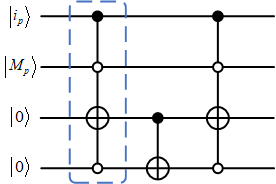}
\caption{The quantum circuit to implement CMP. $i_p\;(p=0,1,...,m-1)$ and $M_p\;(p=0,1,...,m-1)$ are the single binary bit of $(i)_{10}=(i_{m-1}i_{m-2}...i_1i_0)_2$ and $(M)_{10}=(M_{m-1}M_{m-2}...M_1M_0)_2$.}
\label{fig:CMP}
\end{figure}

The CMP operation can be implemented by a quantum circuit shown in Fig.~\ref{fig:CMP}, which is composed of three controlled NOT (CNOT) gates and four inverted controlled NOT (ICNOT) gates. The CNOT gate flips the controlled qubit if its control qubit is $|1\rangle$, while the ICNOT gate flips the controlled qubit if the control qubit is $|0\rangle$. By properly combining these quantum control gates, the controlled qubits can be flipped under certain conditions. For example, after passing the first combined control gate (blue dashed box shown in Fig.~\ref{fig:CMP}), the third qubit will be flipped when the other three control qubits are $|100\rangle$. To implement CMP operation, the quantum circuit of Fig.~\ref{fig:CMP} needs to perform $m$ times, where $m = \lceil {\rm log}_2(M + 1)\rceil$.

In what follows, we present the derivation of preparing the quantum state $\frac{1}{\sqrt{M}}\sum_{j=1}^M|j\rangle$. Assuming there are $n\;(0\le n<2^{m-1})$ qubits' binary value larger than $M$, so we have $M=2^m-n-1$. The states of qubits in different phases (P1-P4) and registers (R1-R4) are presented in Table \ref{tab:QT}. 
Let us start the derivation with the input quantum state $|0^{\otimes m}M00\rangle$.
At first, a $H^{\otimes m}$ gate is applied to R1, the input quantum state is therefore transformed to
\begin{equation}
\frac{1}{\sqrt{2^m}}(|0M00\rangle+|1M00\rangle+...+|(2^m-1)M00\rangle).
\end{equation}
Then, a ICNOT gate is applied to R1 and R4, so that the resultant state can be expressed by
\begin{equation}
\frac{1}{\sqrt{2^m}}(|0M01\rangle+|1M00\rangle+|2M00\rangle+...+|(2^m-1)M00\rangle).
\end{equation}
After that, the CMP operation is applied to R1, R2 and R3, the quantum state is finally transformed to
\begin{equation}\label{finalstate}
\begin{aligned}
&\frac{1}{\sqrt{2^m}}(|0M01\rangle+|1M00\rangle+|2M00\rangle+...+|MM00\rangle+\\
&|(M+1)M10\rangle+...+|(2^m-1)M10\rangle).
\end{aligned}
\end{equation}
From Eq.~(\ref{finalstate}), it is easily to find that the probability of measurement outcome of (0,0) is $\frac{2^m-n-1}{2^m}=\frac{M}{2^m}$. Therefore, the quantum state
\begin{equation}
\frac{1}{\sqrt{M}}(|1\rangle+|2\rangle+...+|M\rangle)=\frac{1}{\sqrt{M}}\sum_{j=1}^M|j\rangle
\end{equation}
can be obtained with the probability $\frac{M}{2^m}$ from R1.

Similarly, we can prepare the quantum state $\frac{1}{\sqrt{U}}\sum_{i=1}^U|i\rangle$ by replace the input qubit $|M\rangle$ with $|U\rangle$. Then, the initial state
\begin{equation}
|\tau_{init}\rangle=\frac{1}{\sqrt{M}}\sum_{j=1}^M|j\rangle \frac{1}{\sqrt{U}}\sum_{i=1}^U|i\rangle|0\rangle|0\rangle
\end{equation}
can be prepared.

\begin{table*}
\newcommand{\tabincell}[2]{\begin{tabular}{@{}#1@{}}#2\end{tabular}}
\centering
\caption{The states of qubits passing quantum circuit shown in Fig.~\ref{fig:inintialstate}.}
\begin{ruledtabular}\label{tab:QT}
\begin{tabular}{ccccc}
 & P1& P2& P3& P4\\
\hline
R1& $|0^{\otimes m}\rangle$ & \tabincell{c}{$\frac{1}{\sqrt{2^m}}(|0\rangle+|1\rangle)^{\otimes m}$\\$=\frac{1}{\sqrt{2^m}}(|0\rangle+|1\rangle+...+|2^m-1\rangle)$} & $\frac{1}{\sqrt{2^m}}(|0\rangle+|1\rangle+...+|2^m-1\rangle)$ & $\frac{1}{\sqrt{2^m}}(|0\rangle+|1\rangle+...+|2^m-1\rangle)$\\

R2& $|M\rangle$ & $|M\rangle$ & $|M\rangle$ & $|M\rangle$\\

R3& $|0\rangle$ & $|0\rangle$ & $|0\rangle$ & $\frac{1}{\sqrt{2^m}}(\underbrace{|0\rangle+...+|0\rangle}_{2^m-n}+\underbrace{|1\rangle+...+|1\rangle}_{n})$\\

R4& $|0\rangle$ & $|0\rangle$ & $\frac{1}{\sqrt{2^m}}(|1\rangle+\underbrace{|0\rangle+...+|0\rangle}_{2^m-1})$ & $\frac{1}{\sqrt{2^m}}(|1\rangle+\underbrace{|0\rangle+...+|0\rangle}_{2^m-1})$
\end{tabular}
\end{ruledtabular}
\end{table*}

\section{Derivation of the probability of measuring the top qubit of c-SWAP test}\label{P0}

As shown in Fig.~\ref{fig:SWAP}, the top qubit $|0\rangle$ is firstly transformed to $\frac{1}{\sqrt{2}}(|0\rangle+|1\rangle)$ by a Hadamard gate. After passing the SWAP gate, the whole quantum system turns to $\frac{1}{\sqrt{2}}(|0\rangle|\rho\rangle|\tau_j\rangle+|1\rangle|\tau_j\rangle|\rho\rangle)$. The top qubit subsequently operated by another Hadamard gate, resulting in the final quantum state
\begin{equation}
|\gamma_j\rangle=\frac{1}{2}|0\rangle(|\rho\rangle|\tau_j\rangle+|\tau_j\rangle|\rho\rangle)+\frac{1}{2}|1\rangle(|\rho\rangle|\tau_j\rangle-|\tau_j\rangle|\rho\rangle).
\end{equation}
If we measure the top qubit of quantum state $|\gamma_j\rangle$, the probability of outcome of $0$ is $P_j(0)=(1+|\langle\rho|\tau_j\rangle|^2)/2$. The detailed derivation is as follows.

Let $\{M_i\}=\{M_0,M_1\}$ be measurement operator set, where
\begin{equation}
\begin{aligned}
M_0=|0\rangle\langle0|\quad M_0^{\dagger}M_0=M_0,\\
M_1=|1\rangle\langle1|\quad M_1^{\dagger}M_1=M_1,
\end{aligned}
\end{equation} 
and $\{M_i\}$ obeys 
\begin{equation}
\begin{aligned}
\sum_i M_i^{\dagger}M_i &=M_0^{\dagger}M_0+M_1^{\dagger}M_1\\
&=M_0+M_1=I,
\end{aligned}
\end{equation}
where $I$ is identity matrix. $P_j(0)$ can be derived by
\begin{widetext} 
\begin{equation}
\begin{aligned}
P_j(0) &=\langle\gamma_j|M_0^{\dagger}M_0|\gamma_j\rangle\\
&=\langle\gamma_j|M_0|\gamma_j\rangle\\
&=\langle\gamma_j||0\rangle\langle0||\gamma_j\rangle\\
&=[\frac{1}{2}(\langle\tau_j|\langle\rho|+\langle\rho|\langle\tau_j|)\langle0|+\frac{1}{2}(\langle\tau_j|\langle\rho|-\langle\rho|\langle\tau_j|)\langle1|]|0\rangle\langle0|[\frac{1}{2}|0\rangle(|\rho\rangle|\tau_j\rangle+\rangle|\tau_j\rangle|\rho\rangle)+\frac{1}{2}|1\rangle(|\rho\rangle|\tau_j\rangle-|\tau_j\rangle|\rho\rangle)]\\
&=[\frac{1}{2}(\langle\tau_j|\langle\rho|+\langle\rho|\langle\tau_j|)\langle0|]|0\rangle\langle0|[\frac{1}{2}|0\rangle(|\rho\rangle|\tau_j\rangle+\rangle|\tau_j\rangle|\rho\rangle)]\\
&=[\frac{1}{2}(\langle\tau_j|\langle\rho|+\langle\rho|\langle\tau_j|)]\langle0|0\rangle\langle0|0\rangle|[\frac{1}{2}(|\rho\rangle|\tau_j\rangle+|\tau_j\rangle|\rho\rangle)]\\
&=[\frac{1}{2}(\langle\tau_j|\langle\rho|+\langle\rho|\langle\tau_j|)][\frac{1}{2}(|\rho\rangle|\tau_j\rangle+|\tau_j\rangle|\rho\rangle)]\\
&=\frac{1}{4}(\langle\tau_j|\langle\rho||\rho\rangle|\tau_j\rangle+\langle\rho|\langle\tau_j||\rho\rangle|\tau_j\rangle+\langle\tau_j|\langle\rho||\tau_j\rangle|\rho\rangle+\langle\rho|\langle\tau_j||\tau_j\rangle|\rho\rangle)\\
&=\frac{1}{4}(2+2|\langle\rho|\tau_j\rangle|^2)\\
&=\frac{1}{2}(1+|\langle\rho|\tau_j\rangle|^2),
\end{aligned}
\end{equation}
\end{widetext} 
where $\langle\rho|\rho\rangle=1$ and $\langle\tau_j|\tau_j\rangle=1$.

\section{Derivation for obtaining quantum state $|\sigma\rangle$ by amplitude estimation}\label{AE}

After $M$ times c-SWAP test, Bob now holds a superposition state $|\gamma\rangle$ (Eq.~(\ref{equ:gamma}) in the main text) whose amplitudes contain all similarities between all known DMCSs and the unknown DMCS. The aim of amplitude estimation is to store these similarities from amplitudes to a quantum bit string, which is convenient for the processing of follow-up quantum search algorithm.

In fact, the superposition state $|\gamma\rangle$ can be decomposed as $|\gamma\rangle=|\gamma_g\rangle+|\gamma_b\rangle$, where $|\gamma_g\rangle=\frac{1}{\sqrt{M}}\sum_{j=1}^M|j\rangle\sqrt{P_j(0)}|0\rangle$ denotes good state that we needed and $|\gamma_b\rangle=\frac{1}{\sqrt{M}}\sum_{j=1}^M|j\rangle\sqrt{1-P_j(0)}|1\rangle$ denotes bad state that we do not needed. Let $a=\langle\gamma_g|\gamma_g\rangle=\frac{1}{M}\sum_{j=1}^MP_j(0)$ denotes the probability that measuring $|\gamma\rangle$ produces a good state, amplitude amplification is first performed to the state $|\gamma\rangle$ by repeatedly applying the unitary operator $Q=-{\mathcal A}S_0{\mathcal A}^{\dagger}S_{\chi}$, where ${\mathcal A}S_0{\mathcal A}^{\dagger}=I-2|\gamma\rangle\langle \gamma|$ and $-S_{\chi}=I-\frac{2}{1-a}|\gamma_b\rangle\langle \gamma_b|$ \cite{Brassard_2002}. We therefore can derive that
\begin{equation}\label{equ:Qg}
\begin{aligned}
Q|\gamma_g\rangle&=(1-2a)|\gamma_g\rangle-2a|\gamma_b\rangle,\\
Q|\gamma_b\rangle&=2(1-a)|\gamma_g\rangle+(1-2a)|\gamma_b\rangle.
\end{aligned}
\end{equation}
From Eqs. (\ref{equ:Qg}), the $Q$ operator can be further expressed as a matrix form
\begin{equation}
Q=\left[ \begin{array}{cc}
1-2a & -2\sqrt{a(1-a)}\\
2\sqrt{a(1-a)} & 1-2a
\end{array}
\right ].
\end{equation}
Assuming ${\rm sin}^2\theta_a=a=\langle\gamma_g|\gamma_g\rangle$, the matrix $Q$ can be rewritten as
\begin{equation}
Q=\left[ \begin{array}{cc}
{\rm cos}(2\theta_a) & -{\rm sin}(2\theta_a)\\
{\rm sin}(2\theta_a) & {\rm cos}(2\theta_a)
\end{array}
\right ],
\end{equation}
so we have
\begin{equation}
\begin{aligned}
Q&={\rm cos}(2\theta_a)I-\iota{\rm sin}(2\theta_a)Y\\
&=e^{-2\iota\theta_a Y},
\end{aligned}
\end{equation}
where $\iota=\sqrt{-1}$ and $Y$ is Pauli Y operator that defined as
\begin{equation}
Y=\left[ \begin{array}{cc}
0 & -\iota\\
\iota & 0
\end{array}
\right ].
\end{equation}
Therefore, we can derive that the eigenvalues of $Q$ are $e^{\pm \iota2\theta_a}$ and the eigenstates of $Q$ are
\begin{equation}
\begin{aligned}
|\Phi_+\rangle&=\frac{1}{\sqrt{2}}(\frac{|\gamma_b\rangle}{\sqrt{1-a}}+\iota\frac{|\gamma_g\rangle}{\sqrt{a}}),\\
|\Phi_-\rangle&=\frac{1}{\sqrt{2}}(\frac{|\gamma_b\rangle}{\sqrt{1-a}}-\iota\frac{|\gamma_g\rangle}{\sqrt{a}}),
\end{aligned}
\end{equation}
so that the quantum state $|\gamma\rangle$ can be expanded on the eigenstates basis of $Q$ as
\begin{equation}
\begin{aligned}
|\gamma\rangle&=|\gamma_b\rangle+|\gamma_g\rangle\\
&=\sqrt{1-a}\frac{1}{\sqrt{2}}(|\Phi_+\rangle+|\Phi_-\rangle)+\iota\sqrt{a}\frac{1}{\sqrt{2}}(|\Phi_+\rangle-|\Phi_-\rangle)\\
&=\frac{1}{\sqrt{2}}(e^{\iota\theta_a}|\Phi_+\rangle+e^{-\iota\theta_a}|\Phi_-\rangle).
\end{aligned}
\end{equation}
As shown in Fig.~\ref{fig:AE}, the Hadamard gates are first applied to $m$ $|0\rangle$ qubits, so that a quantum state $\frac{1}{\sqrt{M}}\sum_{j=0}^{M-1}|j\rangle$ can be produced. Then the operator $Q^j$ is controlled by this state to apply $j$ times of operator $Q$ to quantum state $|\gamma\rangle$, the resultant state is transformed to
\begin{widetext} 
\begin{equation}\label{equ:Qj}
\begin{aligned}
\frac{1}{\sqrt{M}}\sum_{j=0}^{M-1}|j\rangle Q^j|\gamma\rangle &= \frac{1}{\sqrt{M}}\sum_{j=0}^{M-1}|j\rangle[\frac{1}{\sqrt{2}}e^{\iota(2j+1)\theta_a}|\Phi_+\rangle+\frac{1}{\sqrt{2}}e^{-\iota(2j+1)\theta_a}|\Phi_-\rangle]\\
&= \frac{1}{\sqrt{2}}[\frac{1}{\sqrt{M}}\sum_{j=0}^{M-1}e^{\iota2\theta_aj}|j\rangle]e^{\iota\theta_a}|\Phi_+\rangle+\frac{1}{\sqrt{2}}[\frac{1}{\sqrt{M}}\sum_{j=0}^{M-1}e^{-\iota2\theta_aj}|j\rangle]e^{-\iota\theta_a}|\Phi_-\rangle.
\end{aligned}
\end{equation}
\end{widetext} 

After amplitude amplification, phase estimation is subsequently used for estimating the phase information $\theta_a$ of the resultant state. Thus, after applying IQFT ${\mathcal F}_M^{-1}$ on register R1, the quantum state is transformed to
\begin{equation}
\begin{aligned}
|\Phi\rangle=\frac{1}{\sqrt{2}}\left|M\frac{\widetilde{\theta}_a}{\pi}\right\rangle
e^{\iota\theta_a}|\Phi_+\rangle+\frac{1}{\sqrt{2}}\left|M(1-\frac{\widetilde{\theta}_a}{\pi})\right\rangle
e^{-\iota\theta_a}|\Phi_-\rangle,
\end{aligned}
\end{equation}
where $\widetilde{\theta}_a$ is the estimate of $\theta_a$. Let us define $\sigma_1=M\frac{\widetilde{\theta}_a}{\pi}$ and $\sigma_2=M(1-\frac{\widetilde{\theta}_a}{\pi})$, so that ${\rm sin}^2(\widetilde{\theta}_a)={\rm sin}^2(\pi\frac{\sigma_1}{M})$ and ${\rm sin}^2(\widetilde{\theta}_a)={\rm sin}^2(\pi-\pi\frac{\sigma_2}{M})={\rm sin}^2(\pi\frac{\sigma_2}{M})$. Thus, ${\rm sin}^2(\pi\frac{\sigma_1}{M})={\rm sin}^2(\pi\frac{\sigma_2}{M})$, so we have $\sigma_1=\sigma_2=\sigma$. Therefore, the state $|\Phi\rangle$ can be rewritten as
\begin{equation}\label{equ:phi}
\begin{aligned}
|\Phi\rangle=\frac{1}{\sqrt{2}}|\sigma\rangle e^{\iota\theta_a}|\Phi_+\rangle+\frac{1}{\sqrt{2}}|\sigma\rangle e^{-\iota\theta_a}|\Phi_-\rangle,
\end{aligned}
\end{equation}
where 
\begin{equation}
\begin{aligned}
|\sigma\rangle&=\frac{1}{\sqrt{M}}\sum_{j=1}^M|j\rangle \left|\frac{M}{\pi}{\rm arcsin}(\sqrt{\widetilde{P_j}(0)})\right\rangle\\
&=\frac{1}{\sqrt{M}}\sum_{j=1}^M|j\rangle \left|{\rm Sim}_j\right\rangle.
\end{aligned}
\end{equation}
As a result, the similarities between all known DMCSs and the unknown DMCS are stored as a qubit string after amplitude estimation. Note that the quantum state $|\sigma\rangle$ can always be obtained from R1. 

\section{Grover iteration of Grover's algorithm}\label{Grover}

As we mentioned in the main text, the Grover's algorithm includes a number of Grover iterations and a final measurement. The Grover iteration is defined as
\begin{equation}
G_F = {\mathcal D}S_F = W'S_0W'^{\dagger}S_F,
\end{equation}
where $W'|0\rangle=\frac{1}{\sqrt{M}}\sum_{j=1}^{M}|j\rangle$ and $S_0=2|0^m\rangle\langle0^m|-I_m$ \cite{boyer1998tight}.
Assuming there are $t$ different values of $j$ satisfying $F(j)=1$, the initialized quantum state can be expressed as
\begin{equation}\label{D5}
|\Psi_0\rangle=|\Psi(q_0,s_0)\rangle=\sum_{F(j)=1}q_0|j\rangle+\sum_{F(j)\neq1}s_0|j\rangle,
\end{equation}
where $tq^2+(M-t)s^2=1$ and $q_0=s_0=\frac{1}{\sqrt{M}}$. After applying the Grover operator $G_F$ to the $|\Psi(q_0,s_0)\rangle$, it is transformed to
\begin{equation}
\begin{aligned}
|\Psi_1\rangle&=|\Psi(q_1,s_1)\rangle\\
&=|\Psi(\frac{M-2t}{M}q_0+\frac{2(M-t)}{M}s_0,\frac{M-2t}{M}s_0-\frac{2t}{M}q_0)\rangle.
\end{aligned}
\end{equation}
Extending to the general situation, the quantum state $|\Psi_l\rangle=|\Psi(q_l,s_l)\rangle$ can be obtained after $l$ iterations, where
\begin{equation}
\begin{aligned}
q_{l}&=\frac{M-2t}{M}q_{l-1}+\frac{2(M-t)}{M}s_{l-1},\\
s_{l}&=\frac{M-2t}{M}s_{l-1}-\frac{2t}{M}q_{l-1}.
\end{aligned}
\end{equation}
Let ${\rm sin}^2\theta =t/M$ where $\theta \in (0,\pi/2]$, we have 
\begin{equation}
\begin{aligned}
q_l&=\frac{1}{\sqrt{t}}{\rm sin}[(2l+1)\theta],\\
s_l&=\frac{1}{\sqrt{M-t}}{\rm cos}[(2l+1)\theta],
\end{aligned}
\end{equation}
which can be verified by mathematical induction.

The task of the Grover iterations is to continuously amplify the amplitude $q$, so that the probability of obtaining $\sum_{F(j)=1}|j\rangle$ is infinitely close to 1, i.e., we have the following equation
\begin{equation}\label{equ:e6}
\begin{aligned}
t*q_l^2&=t*\frac{1}{t}{\rm sin}^2[(2l+1)\theta]\\
&={\rm sin}^2[(2l+1)\theta]\\
&=1.
\end{aligned}
\end{equation}
From Eq.~(\ref{equ:e6}), we can derive that
\begin{equation}
\begin{aligned}
&{\rm sin}[(2l+1)\theta]=1 \\
\Rightarrow&\;(2l+1)\theta=\frac{\pi}{2} \\
\Rightarrow &\;l=\frac{\pi-2\theta}{4\theta}. 
\end{aligned}
\end{equation}
Since $l$ can only be an integer, thus we have $l=\lfloor \frac{\pi}{4\theta} \rfloor$. In addition, we have $\theta \approx {\rm sin}\theta=\sqrt{\frac{t}{M}}$ when $M$ is large enough. Therefore, it needs $\frac{\pi}{4}\sqrt{\frac{M}{t}}$ Grover iterations to find the target index $j'$.

\section{Calculation of secret key rate for DM CVQKD protocol}\label{DMCVQKD}

In this section, we present the calculation of secret key rate for DM CVQKD protocol with SDP method \cite{Denys2021explicitasymptotic}.
As presented in Sec.~\ref{4.3}, the asymptotic secret key rate of CVQKD protocol with reverse reconciliation and heterodyne detection under collective attack obeys
\begin{equation}
\begin{aligned}
K_{\rm asym} = \beta I_{\rm AB}-\chi_{\rm BE}.
\end{aligned}
\end{equation}

In the heterodyne detection case, the Shannon mutual information between Alice and Bob $I_{\rm AB}$ is given by
\begin{equation}
\begin{aligned}
I_{\rm AB} = {\rm log}_2\frac{V+\chi_{\rm tot}}{1+\chi_{\rm tot}},\label{Iab}
\end{aligned}
\end{equation}
where $V=V_{\rm m}+1$ is the variance of one half of a two-mode squeezed vacuum state, $V_{\rm m}$ is the modulation variance of Alice, $\chi_{\rm tot}=\xi-1+2(1+v_{\rm el})/(\eta T)$ is the total noise referred to the channel input, $T$ is the transmission efficiency, $\xi$ is the excess noise, $\eta$ is efficiency of the detector and $v_{\rm el}$ is noise due to detector electronics.

The Holevo bound of mutual information between Eve and Bob $\chi_{\rm BE}$ is given by
\begin{equation}
\chi_{\rm BE} = \sum_{i=1}^2 G(\frac{\lambda_i-1}{2})-\sum_{i=3}^5 G(\frac{\lambda_i-1}{2}),
\end{equation}
where $G(x) = (x+1){\rm log}_2(x+1)-x{\rm log}_2x$, and $\lambda_i(i=1,2,3,4,5)$ are symplectic eigenvalues of states' covariance matrices, which is given by
\begin{equation}
\begin{aligned}
\lambda_{1,2}=\frac{1}{2}[A\pm\sqrt{A^2-4B}], 
\end{aligned}
\end{equation}
with
\begin{equation}
\begin{aligned}
A&=V^2+T^2(V+\chi_{\rm line})^2-2TZ^2,\\
B&=T^2(V^2+V\chi_{\rm line}-Z^2)^2;
\end{aligned}
\end{equation}
and
\begin{equation}
\begin{aligned}
\lambda_{3,4}=\frac{1}{2}[C\pm\sqrt{C^2-4D}],
\end{aligned}
\end{equation}
with
\begin{equation}
\begin{aligned}
C&=\frac{1}{(T(V+\chi_{\rm tot}))^2}(A\chi_{\rm het}^2+B+1\\
&+2\chi_{\rm het}(V\sqrt{B}+T(V+\chi_{\rm line}))+2TZ^2),\\
D&=(\frac{V+\sqrt{B}\chi_{\rm het}}{T(V+\chi_{\rm tot})})^2;
\end{aligned}
\end{equation}
and the last symplectic eigenvalue $\lambda_5=1$. In the above equations, $\chi_{\rm line}=1/T-1+\xi$ is the channel-added noise, $\chi_{\rm het}=[1+(1-\eta)+2v_{el}]/\eta$ is the detection-added noise, and $Z$ is the correlation between Alice and Bob. In SDP method, $Z$ is defined as
\begin{equation}
\begin{aligned}
Z=2\sqrt{T}{\rm tr}(\tau^{1/2}\hat{a}\tau^{-1/2}\hat{a}^{\dagger})-\sqrt{2T\xi\omega},
\end{aligned}
\end{equation}
where $\hat{a}$ is the annihilation operator, $\hat{a}^{\dagger}$ is the creation operator, and 
\begin{equation}
w=\sum_k p_k(\langle \alpha_k|\hat{a}^{\dagger}_{\tau}\hat{a}_{\tau}|\alpha_k\rangle-|\langle \alpha_k|\hat{a}_{\tau}|\alpha_k\rangle|^2),
\end{equation}
where $\hat{a}_{\tau}=\tau^{1/2}\hat{a}\tau^{-1/2}$, and for an $N$-PSK modulation, 
\begin{equation}
\tau=\frac{1}{N}\sum_{k=0}^{N-1}|\alpha e^{\iota k\theta}\rangle\langle\alpha e^{\iota k\theta}|,
\end{equation}
with $\theta=2\pi/N$.

\end{appendix}

\bibliographystyle{apsrev4-1}
\bibliography{mybibfile.bib}

\end{document}